\documentclass[aps,pre,reprint,showpacs]{revtex4-1}

\usepackage{graphicx}
\usepackage{dcolumn}
\usepackage{bm}

\usepackage{calc}  
\usepackage{amsmath}

\usepackage[utf8]{inputenc}   
\usepackage[T1]{fontenc}      
\usepackage{textcomp}

\begin{document}


\title{Shaping Microwave Fields using Non-Linear Unsolicited Feedback:\\Application to Enhanced Energy Harvesting}

\author{Philipp del Hougne}
\author{Mathias Fink}%
\author{Geoffroy Lerosey}%
\affiliation{Institut Langevin, CNRS UMR 7587, ESPCI Paris, PSL Research University, 1 rue Jussieu, 75005 Paris, France}%

\date{\today}

\begin{abstract}
Wavefront Shaping has emerged over the past decade as a powerful tool to control wave propagation through complex media, initially in optics and more recently also in the microwave domain with important applications in telecommunication, imaging and energy transfer. The crux of implementing wavefront shaping concepts in real life is often its need for (direct) feedback, requiring access to the target to focus on. Here, we present the shaping of a microwave field based on indirect, unsolicited and blind feedback which may be the pivotal step towards practical implementations. With the example of a radiofrequency harvester in a metallic cavity, we demonstrate tenfold enhancement of the harvested power by wavefront shaping based on non-linear signals detected at an arbitrary position away from the harvesting device.

\end{abstract}

\pacs{84.40.-x, 41.20.Jb, 42.25.Dd}

\maketitle


As a wave propagates through a complex medium, its initial wavefront is completely scrambled due to multiple scattering and reflection events occurring inside the medium \cite{NatPhotReview}. Depending on the wave type, very different environments may be considered complex; a thin layer of paint, biological tissue or a multimode fiber at optical wavelengths and cities or disordered cavities for microwaves are common examples \cite{NatPhotReview,choi2015WSforBioMed,cizmar_biomed,choi_biomed,huicaofibres,MIMO_PhysToday}. Formerly, this complete scrambling was perceived as absolutely detrimental to information transfer which in turn is crucial for imaging and communication applications. More recently, various novel techniques emerged that embrace the secondary sources offered by complex media rather than considering them an obstacle. Around the turn of the millennium, the information capacity achievable with MIMO communication systems in complex media was shown to outperform that of free space \cite{MIMO_Science,MIMO_PhysToday} and Time Reversal was developed in acoustics and then also for microwaves \cite{TR_fink,EMTR_prl}. A bit later, wavefront shaping was introduced in optics \cite{mosk_SLM}.

Since then, wavefront shaping in complex media has enabled fascinating demonstrations such as focusing beyond the Rayleigh limit \cite{choi2011subwavelengthFOC,Mosk_vis_subwavelength_foc_byWS,park2013subwavelength,mosk_disorder4perfectFOC}, the spatiotemporal refocusing of distorted pulses \cite{aulbach_STF,katz_STF,publikation3} and sub-sampled compressive imaging \cite{liutkus14}, to name a few. Furthermore, measuring the complex medium's transmission matrix \cite{popoff_prl,popoff_NatComm,LD_binDMD,park2013largeTM,mikael_STF,TM_RevMed} provided information about important statistical properties and the transmission eigenchannels of the medium \cite{EigChan_Choi_NPhot,Mosk_WS_OptTrans,pena2014single}, as well as being an open-loop tool in contrast to iterative focusing algorithms.

However, ten years after Vellekoop and Mosk's first demonstration in optics \cite{mosk_SLM}, focusing by wavefront shaping has not yet become an omnipresent technique in commercial imaging devices, medical therapy or the telecommunication industry. A challenging hurdle on the path from academic proof-of-concepts towards real-life applications is usually the need for a feedback signal from the target point(s) to focus on, a common characteristic of all wavefront shaping techniques. For medical applications, a camera cannot be placed inside the biological tissue and implanting objects that generate fluorescence or harmonics might be too invasive. 
Similarly, improving signal reception on a wireless device in an indoor environment by
wavefront shaping \cite{SMM_PoC} requires real-time access to the device’s received signal strength indicator, which is possible to some extent in WIFI, for instance, but difficult to imagine for low energy IoT devices.

These difficulties with direct feedback motivate the identification of indirect feedback schemes. 
Indirect solicited feedback about the target intensity was already successfully employed in fluorescence experiments in optics \cite{FluoMosk}; indirect unsolicited feedback has been demonstrated with MR-guided ultrasound focusing in acoustics and by exploiting either the photo-acoustic effect or two-photon fluorescence in optics, all using biological tissue \cite{MRfeedback,Gigan_NonInvTM_PA,OriNL}. In this Letter, we transpose this concept of wavefront shaping with indirect unsolicited feedback to the microwave domain. Our target to focus on is a non-linear device, a radio-frequency (RF) harvester, that captures the ambient microwave signal and rectifies it into a DC output. The rectification involving diodes is a non-linear process inevitably generating non-linear signals that are re-emitted and constitute our indirect feedback.

Incidentally, our work addresses a key challenge of current RF harvesting set-ups: the harvested voltages are too low for real-life applications \cite{PowByWifi}. Potentially, RF harvesting is a promising technique in the advent of concepts such as \textit{Smart Home} and \textit{Internet of the Things} (IoT). It may enable the wireless and battery-free powering of many low-power sensors, recycling the energy of the ubiquitous RF fields in our urban environments and thereby constituting a step towards a greener future. 

Using simple electronically reconfigurable reflector arrays, so-called Spatial-Microwave-Modulators (SMMs) \cite{SMM_design}, we create constructive interferences of the reflected waves at the position of the harvester. Thereby we focus the wave field and enhance the harvested energy that depends in a monotonous but non-linear way on the incident field intensity. Firstly, we demonstrate in the controlled environment of a disordered microwave cavity the significant enhancement of the harvester's voltage output by optimizing the incident wavefront, using the harvested voltage as direct feedback. Secondly, we maximize once again the harvested voltage, but this time using an indirect unsolicited feedback: the strength of non-linear signatures detected inside the cavity at an arbitrary location away from the harvester.

We use a disordered metallic cavity ($1.1 \ \mathrm{m}^3$; $Q=835$) 
that constitutes a static, well-controllable complex medium for our proof-of-concept experiments. In the microwave domain, reverberant media are very common: electromagnetic compatibility tests require reverberation chambers \cite{hill_electromagnetic_2009,JB_PRL,JB_PRE}, open disordered cavities currently attract a lot of interest for computational imaging \cite{DavidSmith_CompImag_APL,sleasman2016microwave,SmithNumerical,davidsmithscirep} and indoor environments trap telecommunication signals \cite{MIMO_PhysToday,SMM_PoC}. The SMM covers roughly $7\%$ of the cavity walls with $102$ binary elements whose boundary conditions can be switched dynamically between Dirichlet and Neumann for frequencies within a $100 \ \mathrm{MHz}$ bandwidth around $2.47 \ \mathrm{GHz}$; their working principle is outlined in the inset in Fig.~1 and in Ref.~\cite{SMM_design}. 

\begin{figure}[ht]
	\begin{center}
\includegraphics [width=\columnwidth] {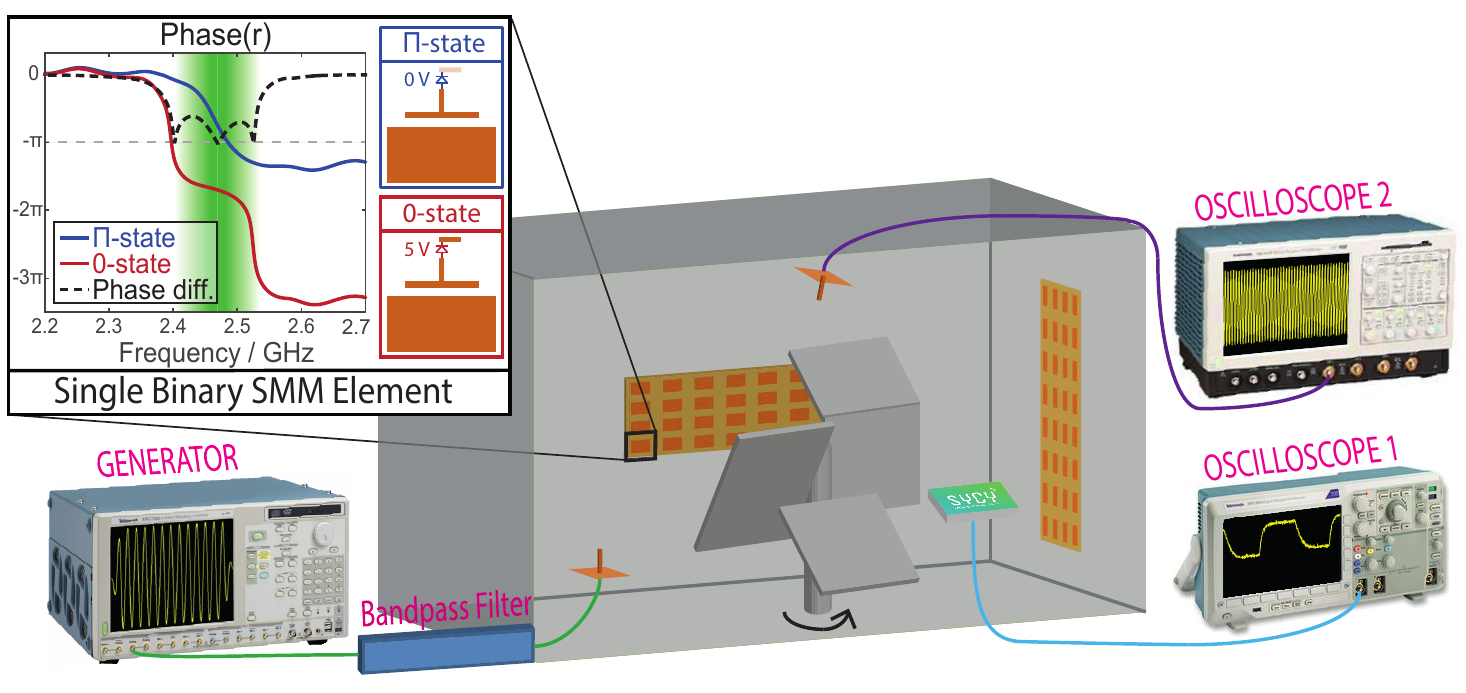}
	\caption{Schematic of experimental set-up. Using a wave generator, a quasi-continuous ambient field is 	generated inside a disordered metallic cavity. This field can be shaped with Spatial-Microwave-Modulators that partially cover the cavity walls. Oscilloscope 1 monitors the voltage output of a radio-frequency harvester. The high sampling rate Oscilloscope 2 is used to analyze the spectrum at an arbitrary position away from the harvester. A mode-stirrer rotation by $12^\circ$ conveniently realizes disorder. Inset adapted from Ref.~\cite{SMM_design}.}
	\label{fig1}
	\end{center}
\end{figure}

With an arbitrary signal generator (sampling at $10 \ \mathrm{GHz}$), we mimic a continuously excited ambient field by emitting a 
$30 \ \mathrm{\mu s} $ long signal within the $2.4 \ \mathrm{GHz}$ WIFI band; a bandpass filter ($2.38-2.52 \ \mathrm{GHz}$) cleans the signal before it is emitted into the cavity by a monopole antenna adapted for WIFI frequencies in free space. 

The RF harvester is a commercial prototype (cf. acknowledgments) that uses a low-power Schottky diode circuit to rectify the captured microwave signal \cite{HarviDesign}. The  results we present stand on their own and are independent of the harvester's detailed operating mode. The employed device harvests most efficiently around $2.42 \ \mathrm{GHz}$. With the low-frequency Oscilloscope 1 ($1 \ \mathrm{M\Omega}$; $100  \ \mathrm{MHz}$; 8  bits), triggered by the generator, we monitor the harvested voltage. As exemplified in Fig.~2(a), it takes a few microseconds for the harvested voltage to rise, and a bit longer to decay after the excitation signal stops. The repetition rate of the generator is chosen such that the cycles do not overlap; over a $20 \ \mathrm{\mu s}$ interval (highlighted in green in Fig.~2(a)), the harvested signal is stationary. In the following, harvested voltage $V_{harv}$ refers to the average signal received during this stable interval.

For the indirect feedback scheme in the second part, a further WIFI monopole antenna is placed at an arbitrary location inside the cavity outside the harvester's line of sight; the high sampling rate Oscilloscope 2 measures the received signal, again triggered by the generator. A $2 \ \mathrm{\mu s}$ interval, sampled at $25 \ \mathrm{GHz}$ and averaged over $50$ measurements, is acquired and then Fourier transformed to quantify the intensity of non-linear signals in the spectrum. 
Using adapted antennas, appropriate filters or lock-in detection would be simple, cheap and well-established means to improve the acquisition robustness and simultaneously remove the need for the costly Oscilloscope 2.
Note that the frequency of optimal operation varies across our employed equipment (monopole antennas, SMM, harvester); 
while this does not hinder the intended proof-of-concept, quantitatively even better results are to be expected with refined equipment.

Work with disordered media usually requires averaging over many realizations of disorder to get a representative idea of the underlying physics. An individual optimization outcome strongly depends on the initial conditions, e.g. whether the speckle-like field initially has a node or anti-node at the target position. We conveniently realize disorder with the mode-stirrer indicated in Fig.~1: rotating it by $12^\circ$ yields a ``new''  disordered cavity with the same global parameters (volume, quality factor, \dots) but a different geometry, enabling a total of 30 independent realizations. As the SMM has a large control over the wave field in this set-up \cite{publikation1}, the experiment can moreover be repeated several times for each mode-stirrer position, starting with a different random SMM configuration each time. Random SMM configurations effectively constitute different cavity geometries preserving the global parameters, too.

\begin{figure}[h]
	\begin{center}
\includegraphics [width=\columnwidth] {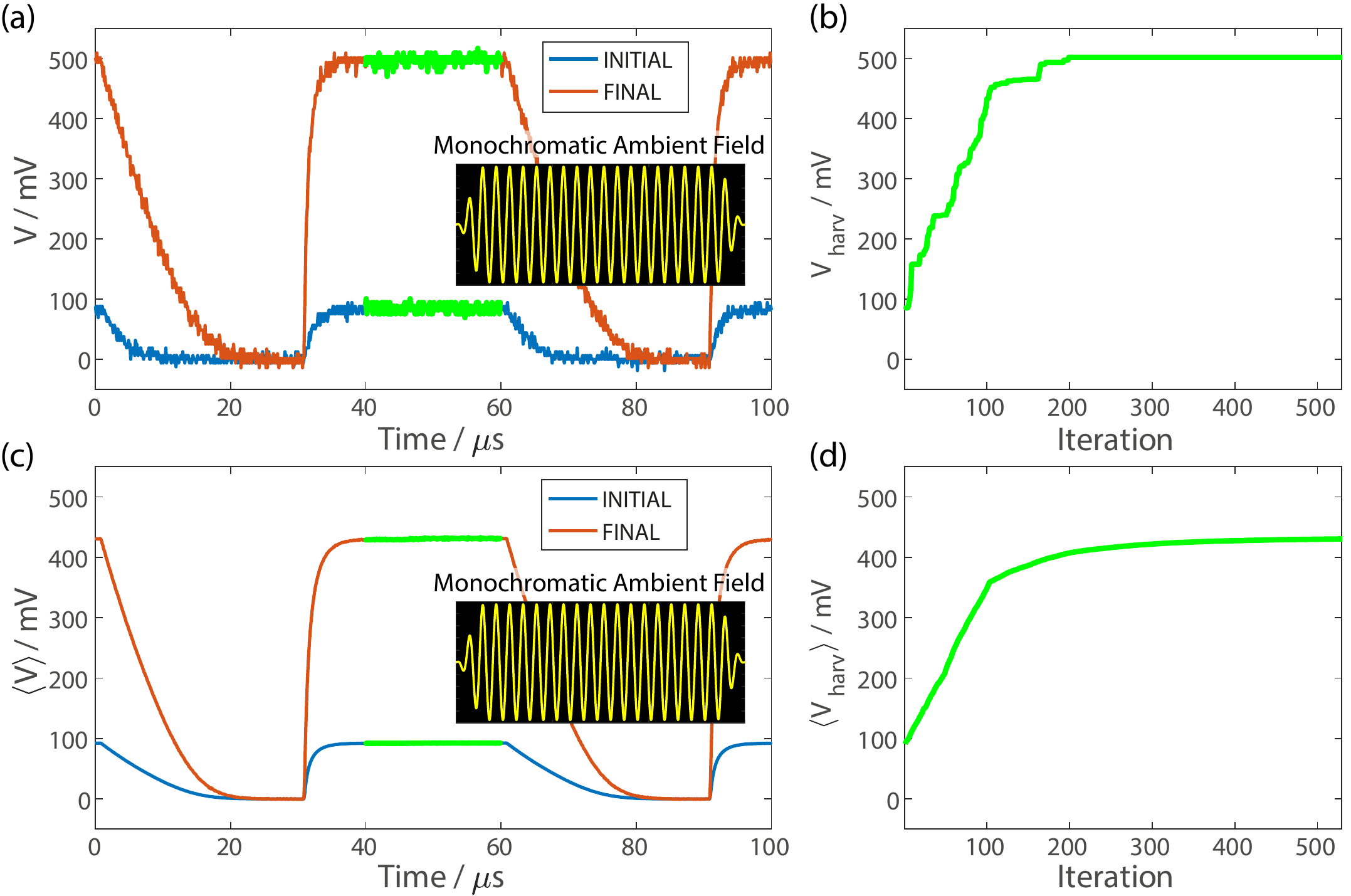}
	\caption{Experimental method exemplified for a monochromatic ambient field using direct feedback. (a) shows the harvester's voltage output, monitored on Oscilloscope 1, before (blue) and after (red) optimization. The interval chosen to estimate $V_{harv}$ is indicated in green, and the evolution of $V_{harv}$ over the course of the iterative optimization is displayed in (b). (c) and (d) show the quantities presented in (a) and (b) averaged over $300$ realizations of disorder.}
	\label{fig2}
	\end{center}
\end{figure}

To begin with, we consider the case of an ambient monochromatic field that we would like to harvest, using the harvested voltage as direct feedback. To identify the optimum SMM configuration, we use an iterative continuous sequential optimization algorithm \cite{moskWSalgo}. Element after element it tests which of the two possible SMM states brings us closer to our objective of maximizing a chosen cost function $CF$, here $CF = V_{harv}$. This procedure is summarized in Fig.~2 where we show the harvester output before and after optimization in (a) and the dynamics of the optimization in (b); (c) and (d) present the same quantities averaged over $300$ realizations of disorder. Note that the number of iterations required until saturation in (b) is about twice the number of SMM pixels. Unlike the first optics experiments that used this iterative method to focus through multiply scattering paint layers, we cannot limit ourselves to testing each element only once; instead we have to retest them several times due to the reverberation that correlates the optimum states of different elements.

Next, we explore how the harvesting enhancement by wavefront shaping with our setup depends on the ambient monochromatic field's frequency and power. The generator's peak-to-peak voltage $V_{pp}$ is used to alter the ambient field's power. Each resulting data point displayed in Fig.~3(a) is the average over $300$ realizations. Here, we chose a representation in terms of voltage (rather than power), as the minimum voltage requirements, even by DC-DC converters, were identified as limiting factor in Ref.~\cite{PowByWifi} for harvesting schemes to be useful in practice. 
Since our employed equipment's frequency responses are not flat, a slight frequency dependence is evident in Fig.~3(a). 
The power dependence may be surprising, since power is not a variable appearing in the theoretical model used to explain traditional monochromatic wavefront shaping experiments in terms of degrees of freedom \cite{publikation1}. This can be understood, however, from the fact that $V_{harv}$ depends in a monotonous but not necessarily linear manner on the ambient monochromatic field's intensity ${\left| S(f_{0},\mathbf{r_{0}}) \right|}^2$ at the harvester's position $\mathbf{r_{0}}$. Here, the mean voltage enhancements vary between about $4$ and $6$, the corresponding power enhancements thus being on the order of $20$ to $30$; the attained enhancement is larger for weaker ambient fields. This power dependence, likely to be generalizable to most harvesting devices, works in favor of our proposal to enhance harvesting by wavefront shaping, in particular in the case of (realistic) weak ambient fields.

\begin{figure}[h]
	\begin{center}
\includegraphics [width=6 cm] {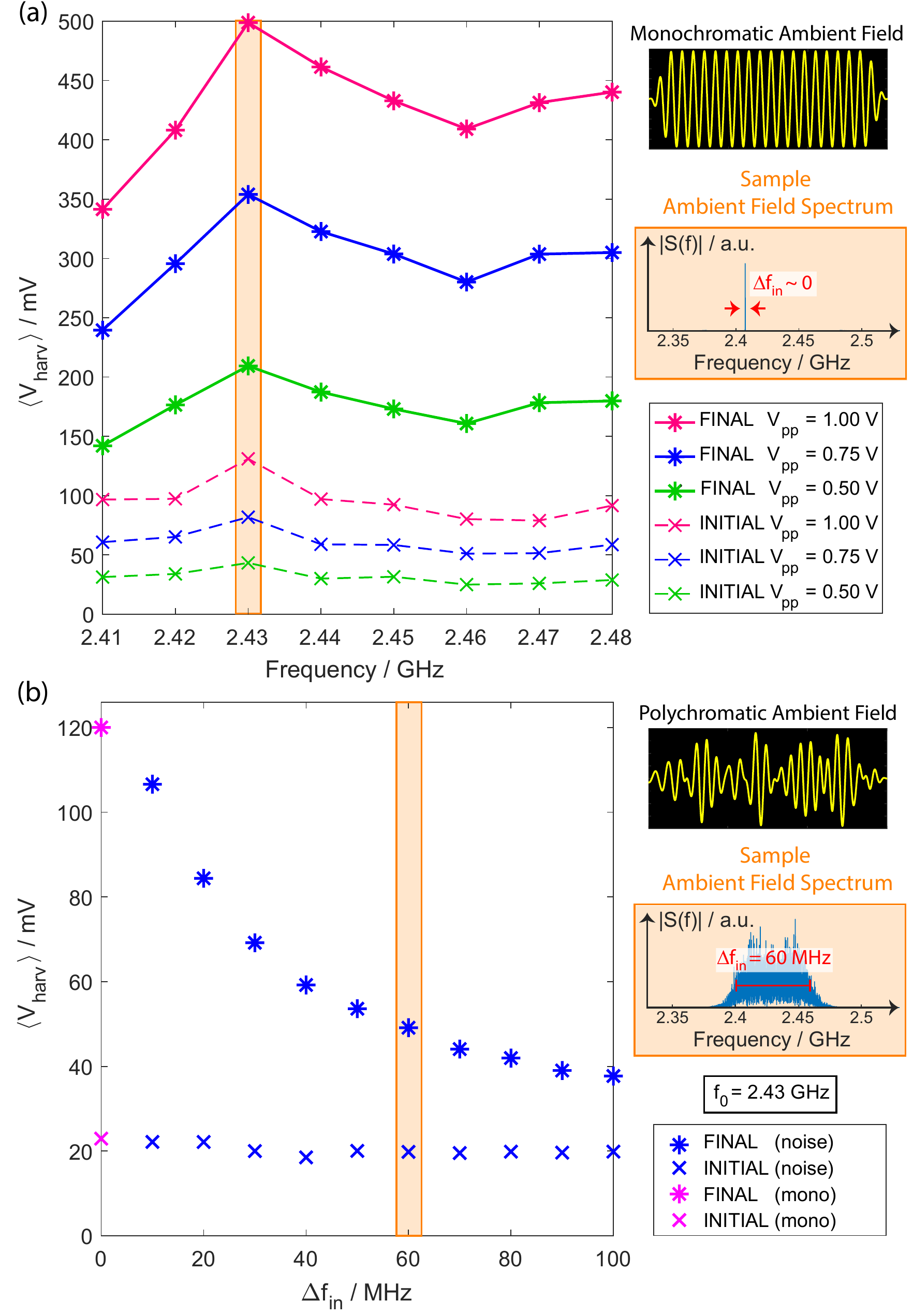}
	\caption{The harvested voltages before and after direct feedback based wavefront shaping, (a) for monochromatic fields of different frequencies and powers (cf. legend), averaged over $300$ realizations of disorder; (b) for polychromatic (noise) fields of different bandwidths $\Delta f_{in}$, centered on $f_0 = 2.43 \ \mathrm{GHz}$, averaged over $150$ realizations.}
	\label{fig3}
	\end{center}
\end{figure}

How well does wavefront shaping based harvesting enhancement do in a more realistic, polychromatic ambient field? To explore this question, we work with noise signals \footnote{Bandpass filtered long series of numbers generated with MatLab's uniform random number generator.}, emitted by the generator, of different bandwidths $\Delta f_{in}$ centered on $2.43 \ \mathrm{GHz}$. We observe a clear decrease of the achievable voltage enhancement from a factor of about $5$ to a factor of about $2$, as $\Delta f_{in}$ is increased. This tendency can be understood with traditional wavefront shaping tools. In the case of a polychromatic ambient field, the harvested voltage is essentially equivalent to incoherent polychromatic focusing with unknown weights $w(f)$ for different frequencies: $V_{harv} \approx \int_{\Delta f_{in}}w(f)\left|S(f)\right|^2 \ \textnormal{d}f$. Wavefront shaping can relocate a certain amount of energy that is on average equally spread across the $1 + \Delta f_{in}/\Delta f_{corr}$ independent frequencies, where $\Delta f_{corr} = f_0 / Q$ is the cavity correlation frequency; the literature contains multiple reports confirming this experimentally \cite{SpectIncoh1,SpectIncoh2,SpectIncoh3,SpectIncoh4,SpectIncoh5}. In Fig.~3(b) the decrease of the attainable enhancement is quite drastic as our highly reverberating cavity has a correlation frequency of a few $\mathrm{MHz}$, implying a high number of independent frequencies inside $\Delta f_{in}$. Yet in lossier and leakier real life systems $\Delta f_{corr}$ would be rather on the order of a few tens of $\mathrm{MHz}$ such that real scenarios would stay within the very upper part of the curve, not experiencing major drawbacks from broadband operation.


Having demonstrated the viability of wavefront shaping to enhance the harvested voltages both in monochromatic and polychromatic ambient fields using direct feedback, we now turn to the indirect feedback case. The diode-based rectifier circuit inside the harvester is intrinsically a source of non-linearities that are re-emitted into the cavity by the harvester's receiving antenna. 
Approximated to first-order, the strength of the non-linear re-emissions increases monotonously as the excitation intensity incident on the harvester, ${\left| S(f_{0},\mathbf{r_{0}}) \right|}^2$, rises. The intensity ${\left| S(f_{NL},\mathbf{r_{NL}}) \right|}^2$ of a non-linear signature of frequency $f_{NL}$ at position $\mathbf{r_{NL}}$ away from the harvester may thus serve as feedback about the excitation intensity incident on the harvester that is unsolicited as it is generated naturally and inevitably. Moreover, it constitutes a blind feedback in the sense that we focus the wave field on the harvester without any knowledge of its position $\mathbf{r_{0}}$ in space.

Under which circumstances will $CF={\left| S(f_{NL},\mathbf{r_{NL}}) \right|}^2$ provide a reliable feedback about ${\left| S(f_{0},\mathbf{r_{0}}) \right|}^2$? Changes in ${\left| S(f_{NL},\mathbf{r_{NL}}) \right|}^2$ must occur \emph{only} in response to changes in ${\left| S(f_{0},\mathbf{r_{0}}) \right|}^2$. If there were sources other than the harvester emitting at $f_{NL}$, the detected magnitude ${\left| S(f_{NL},\mathbf{r_{NL}}) \right|}^2$ of the interference of all those $f_{NL}$-sources would be sensitive to relative phase differences between the sources. Similarly, if the wave field at $f_{NL}$ was modulated by the SMM, the value measured for ${\left| S(f_{NL},\mathbf{r_{NL}}) \right|}^2$ would heavily depend on the SMM state, as well as on ${\left| S(f_{0},\mathbf{r_{0}}) \right|}^2$. Fortunately, neither of those scenarios arises in our setup; otherwise, either or both could be circumvented by working with $\langle {\left| S(f_{NL},\mathbf{r_{NL}}) \right|}^2 \rangle_{\mathrm{independent}\ \mathbf{r_{NL}}}$, the average of ${\left| S(f_{NL},\mathbf{r_{NL}}) \right|}^2$ over several independent positions $\mathbf{r_{NL}}$.

\begin{figure}[h]
	\begin{center}
\includegraphics [width=\columnwidth] {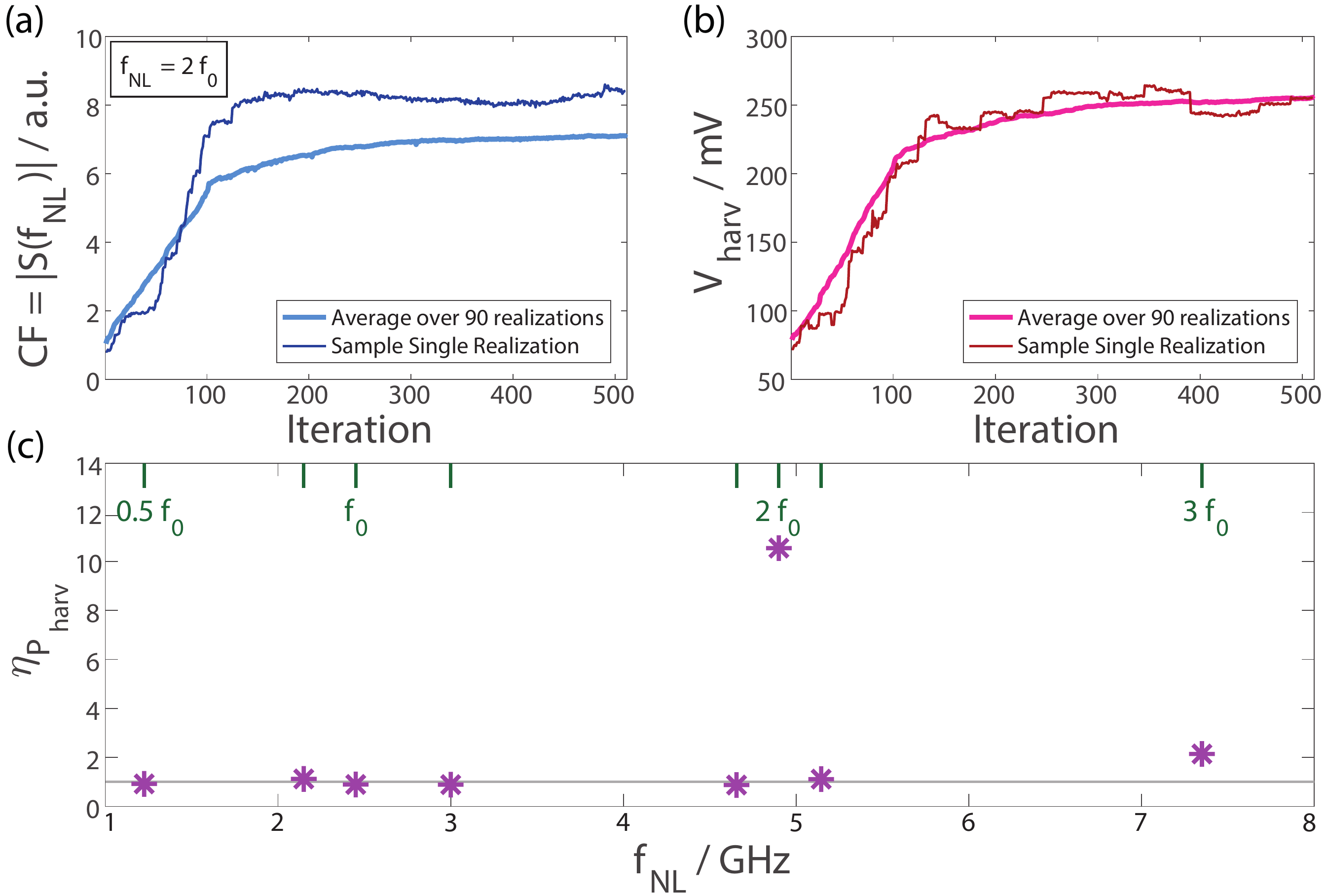}
	\caption{Wavefront Shaping with indirect, unsolicited, blind feedback $CF={\left| S(f_{NL},\mathbf{r_{NL}}) \right|}^2$. For the monochromatic case with $f_{NL}=2f_0$, we display the optimization dynamics of the cost function in (a) and the corresponding harvested voltage in (b), both for a single realization and the average over 90 realizations of disorder. The average enhancement of the harvested power $\eta_{P_{harv}} = \eta_{V_{harv}}^2$ is displayed in (c) for a range of different choices of $f_{NL}$.}
	\label{fig4}
	\end{center}
\end{figure}

To demonstrate the feasibility of the indirect feedback based harvesting enhancement scheme, we here chose to work with a monochromatic ambient field at $2.43 \ \mathrm{GHz}$ and limit ourselves to $90$ realizations, as our set-up is not optimized in terms of speed. In the top row of Fig.~4 we present the example of $f_{NL} = 2f_0$, illustrating both a single realization and the average over 90 realizations of disorder. On the left in Fig.~4(a) we show the evolution of the non-linear feedback signal, over the course of the iterative optimization. On the right in Fig.~4(b) we display how the harvested voltage at the target position is enhanced. Non-linearities being naturally weak in comparison to the excitation signal, the individual realization suffers notably more from noise than in Fig.~2(b) where we used the harvested voltage as direct feedback. 

The achieved mean enhancement of the harvested voltage of $3.3$, albeit being substantial and corresponding to a ten-fold enhancement of the harvested power, is nonetheless notably lower than the results from direct feedback seen in Fig.~3(a). This can of course be attributed in particular to the unfavorable dynamic range of our temporal measurement of ${\left| S(f_{NL},\mathbf{r_{NL}}) \right|}^2$.

Furthermore, we have tested frequencies other than $2f_0$ to provide indirect feedback, the results being on display in Fig.~4(c), in terms of the average enhancement of the harvested power $\eta_{P_{harv}} = \eta_{V_{harv}}^2 = \langle V_{harv}^{FIN} \rangle ^2 / \langle V_{harv}^{INIT} \rangle ^2$. It can be seen that only $2f_0$ and $3f_0$ result in an enhancement of the harvested voltage, which is significantly stronger in the case of $2f_0$. This confirms that, as one might have anticipated intuitively, the best candidate to work with is the second harmonic \cite{AnlageNLTR,AnlageNLprl}. At $f_{NL} = f_0$, the quantity ${\left| S(f_0,\mathbf{r_{NL}})\right|}^2$ has of course been heavily enhanced but this did not correlate at all with the evolution of ${\left| S(f_0,\mathbf{r_{0}})\right|}^2$: the value of ${\left| S(f_0,\mathbf{r_{NL}})\right|}^2$ is dominated by the SMM's state and the emitted excitation signal. The other tested frequencies were arbitrary, thus not corresponding to any non-linear signatures, such that they did not yield any enhancement either. We have also verified that results similar to the ones presented in Fig.~4 are obtained for different $\mathbf{r_{NL}}$ and $f_0$, but they have been omitted for clarity's sake here.


To conclude, in this Letter we have started off by proving that shaping an ambient microwave field in a reverberant medium to concentrate it on a radiofrequency harvester may constitute an innovative improvement to current RF harvesting schemes that typically do not harvest sufficiently high voltages. Using the harvested voltage as direct feedback, we demonstrated significant harvesting enhancements both for a variety of monochromatic and polychromatic wave fields. Then, we exploited the non-linear nature of the diode-based harvesting device that inevitably causes the re-emission of non-linear signatures into the cavity. By measuring the intensity of the second harmonic at an arbitrary position away from the harvester, we obtained an indirect, unsolicited and blind feedback about the ambient field intensity at the harvester. This enabled a tenfold enhancement of the harvested power with our current setup.

Using indirect, unsolicited and blind feedback, removing the need to access the target or its spatial position directly, may be the crucial bridge between an academic concept of focusing by wavefront shaping and its application in practice, in many cases.
We expect our work to be particularly useful for emerging concepts such as \textit{Smart Homes} that envisage to populate homes and factories with many low-power sensors. With the aid of spatial microwave modulators (SMMs), IoT devices and sensors could be powered wirelessly, harvesting the ambient omnipresent RF fields. 
Improving the SMM design \cite{SMM_design,RAcummer,RApc,RAli}, matching the operating bandwidths of SMM and harvester as well as covering more than only $7\%$ of the walls with SMMs should easily enable much higher enhancements than reported here and counterbalance the decrease in wavefront shaping ability of the SMM in less reverberant (lower $Q$) realistic environments \cite{publikation1}. 
Using non-linear feedback is expected to simultaneously compress the impulse response of \textit{pulsed} ambient fields temporally \cite{katz_STF,BretagnePRE} but realistic communication signals typically have time-bandwidth products orders of magnitude above unity \cite{PowByWifi}.
Multi-target focusing \cite{mosk_SLM,TM_RevMed} may require additional techniques like frequency-tagging to avoid focusing only on the target emitting the strongest non-linearity \cite{OriNL}. 
For other applications such as wireless phone charging that require a lot more power than available in the ambient RF fields, an active emission is certainly necessary; employing our indirect feedback approach might be considered in such wireless power transfer \cite{WTP_Smith,Anlage_syntheticTR,AnlageNLapplied} scenarios, too.

\bigskip

We gratefully acknowledge the start-up SYCY (http://www.sycy.fr) for providing us with a prototype of their RF harvester. P.d.H. acknowledges funding from the French ``Ministère de la Défense, Direction Générale de l’Armement''. This work is supported by LABEX WIFI (Laboratory of Excellence within the French Program ``Investments for the Future'') under references ANR-10-LABX-24 and ANR-10-IDEX-0001-02 PSL*.


\begin{thebibliography}{59}%
\makeatletter
\providecommand \@ifxundefined [1]{%
 \@ifx{#1\undefined}
}%
\providecommand \@ifnum [1]{%
 \ifnum #1\expandafter \@firstoftwo
 \else \expandafter \@secondoftwo
 \fi
}%
\providecommand \@ifx [1]{%
 \ifx #1\expandafter \@firstoftwo
 \else \expandafter \@secondoftwo
 \fi
}%
\providecommand \natexlab [1]{#1}%
\providecommand \enquote  [1]{``#1''}%
\providecommand \bibnamefont  [1]{#1}%
\providecommand \bibfnamefont [1]{#1}%
\providecommand \citenamefont [1]{#1}%
\providecommand \href@noop [0]{\@secondoftwo}%
\providecommand \href [0]{\begingroup \@sanitize@url \@href}%
\providecommand \@href[1]{\@@startlink{#1}\@@href}%
\providecommand \@@href[1]{\endgroup#1\@@endlink}%
\providecommand \@sanitize@url [0]{\catcode `\\12\catcode `\$12\catcode
  `\&12\catcode `\#12\catcode `\^12\catcode `\_12\catcode `\%12\relax}%
\providecommand \@@startlink[1]{}%
\providecommand \@@endlink[0]{}%
\providecommand \url  [0]{\begingroup\@sanitize@url \@url }%
\providecommand \@url [1]{\endgroup\@href {#1}{\urlprefix }}%
\providecommand \urlprefix  [0]{URL }%
\providecommand \Eprint [0]{\href }%
\providecommand \doibase [0]{http://dx.doi.org/}%
\providecommand \selectlanguage [0]{\@gobble}%
\providecommand \bibinfo  [0]{\@secondoftwo}%
\providecommand \bibfield  [0]{\@secondoftwo}%
\providecommand \translation [1]{[#1]}%
\providecommand \BibitemOpen [0]{}%
\providecommand \bibitemStop [0]{}%
\providecommand \bibitemNoStop [0]{.\EOS\space}%
\providecommand \EOS [0]{\spacefactor3000\relax}%
\providecommand \BibitemShut  [1]{\csname bibitem#1\endcsname}%
\let\auto@bib@innerbib\@empty
\bibitem [{\citenamefont {Mosk}\ \emph {et~al.}(2012)\citenamefont {Mosk},
  \citenamefont {Lagendijk}, \citenamefont {Lerosey},\ and\ \citenamefont
  {Fink}}]{NatPhotReview}%
  \BibitemOpen
  \bibfield  {author} {\bibinfo {author} {\bibfnamefont {A.~P.}\ \bibnamefont
  {Mosk}}, \bibinfo {author} {\bibfnamefont {A.}~\bibnamefont {Lagendijk}},
  \bibinfo {author} {\bibfnamefont {G.}~\bibnamefont {Lerosey}}, \ and\
  \bibinfo {author} {\bibfnamefont {M.}~\bibnamefont {Fink}},\ }\href@noop {}
  {\bibfield  {journal} {\bibinfo  {journal} {Nat. Photonics}\ }\textbf
  {\bibinfo {volume} {6}},\ \bibinfo {pages} {283} (\bibinfo {year}
  {2012})}\BibitemShut {NoStop}%
\bibitem [{\citenamefont {Kim}\ \emph {et~al.}(2015)\citenamefont {Kim},
  \citenamefont {Choi}, \citenamefont {Choi}, \citenamefont {Yoon},\ and\
  \citenamefont {Choi}}]{choi2015WSforBioMed}%
  \BibitemOpen
  \bibfield  {author} {\bibinfo {author} {\bibfnamefont {M.}~\bibnamefont
  {Kim}}, \bibinfo {author} {\bibfnamefont {W.}~\bibnamefont {Choi}}, \bibinfo
  {author} {\bibfnamefont {Y.}~\bibnamefont {Choi}}, \bibinfo {author}
  {\bibfnamefont {C.}~\bibnamefont {Yoon}}, \ and\ \bibinfo {author}
  {\bibfnamefont {W.}~\bibnamefont {Choi}},\ }\href@noop {} {\bibfield
  {journal} {\bibinfo  {journal} {Opt. Express}\ }\textbf {\bibinfo {volume}
  {23}},\ \bibinfo {pages} {12648} (\bibinfo {year} {2015})}\BibitemShut
  {NoStop}%
\bibitem [{\citenamefont {{\v{C}}i{\v{z}}m{\'a}r}\ and\ \citenamefont
  {Dholakia}(2011)}]{cizmar_biomed}%
  \BibitemOpen
  \bibfield  {author} {\bibinfo {author} {\bibfnamefont {T.}~\bibnamefont
  {{\v{C}}i{\v{z}}m{\'a}r}}\ and\ \bibinfo {author} {\bibfnamefont
  {K.}~\bibnamefont {Dholakia}},\ }\href@noop {} {\bibfield  {journal}
  {\bibinfo  {journal} {Opt. Express}\ }\textbf {\bibinfo {volume} {19}},\
  \bibinfo {pages} {18871} (\bibinfo {year} {2011})}\BibitemShut {NoStop}%
\bibitem [{\citenamefont {Choi}\ \emph {et~al.}(2012)\citenamefont {Choi},
  \citenamefont {Yoon}, \citenamefont {Kim}, \citenamefont {Yang},
  \citenamefont {Fang-Yen}, \citenamefont {Dasari}, \citenamefont {Lee},\ and\
  \citenamefont {Choi}}]{choi_biomed}%
  \BibitemOpen
  \bibfield  {author} {\bibinfo {author} {\bibfnamefont {Y.}~\bibnamefont
  {Choi}}, \bibinfo {author} {\bibfnamefont {C.}~\bibnamefont {Yoon}}, \bibinfo
  {author} {\bibfnamefont {M.}~\bibnamefont {Kim}}, \bibinfo {author}
  {\bibfnamefont {T.~D.}\ \bibnamefont {Yang}}, \bibinfo {author}
  {\bibfnamefont {C.}~\bibnamefont {Fang-Yen}}, \bibinfo {author}
  {\bibfnamefont {R.~R.}\ \bibnamefont {Dasari}}, \bibinfo {author}
  {\bibfnamefont {K.~J.}\ \bibnamefont {Lee}}, \ and\ \bibinfo {author}
  {\bibfnamefont {W.}~\bibnamefont {Choi}},\ }\href@noop {} {\bibfield
  {journal} {\bibinfo  {journal} {Phys. Rev. Lett.}\ }\textbf {\bibinfo
  {volume} {109}},\ \bibinfo {pages} {203901} (\bibinfo {year}
  {2012})}\BibitemShut {NoStop}%
\bibitem [{\citenamefont {Xiong}\ \emph {et~al.}(2017)\citenamefont {Xiong},
  \citenamefont {Ambichl}, \citenamefont {Bromberg}, \citenamefont {Redding},
  \citenamefont {Rotter},\ and\ \citenamefont {Cao}}]{huicaofibres}%
  \BibitemOpen
  \bibfield  {author} {\bibinfo {author} {\bibfnamefont {W.}~\bibnamefont
  {Xiong}}, \bibinfo {author} {\bibfnamefont {P.}~\bibnamefont {Ambichl}},
  \bibinfo {author} {\bibfnamefont {Y.}~\bibnamefont {Bromberg}}, \bibinfo
  {author} {\bibfnamefont {B.}~\bibnamefont {Redding}}, \bibinfo {author}
  {\bibfnamefont {S.}~\bibnamefont {Rotter}}, \ and\ \bibinfo {author}
  {\bibfnamefont {H.}~\bibnamefont {Cao}},\ }\href@noop {} {\bibfield
  {journal} {\bibinfo  {journal} {Opt. Express}\ }\textbf {\bibinfo {volume}
  {25}},\ \bibinfo {pages} {2709} (\bibinfo {year} {2017})}\BibitemShut
  {NoStop}%
\bibitem [{\citenamefont {Simon}\ \emph {et~al.}(2001)\citenamefont {Simon},
  \citenamefont {Moustakas}, \citenamefont {Stoytchev},\ and\ \citenamefont
  {Safar}}]{MIMO_PhysToday}%
  \BibitemOpen
  \bibfield  {author} {\bibinfo {author} {\bibfnamefont {S.~H.}\ \bibnamefont
  {Simon}}, \bibinfo {author} {\bibfnamefont {A.~L.}\ \bibnamefont
  {Moustakas}}, \bibinfo {author} {\bibfnamefont {M.}~\bibnamefont
  {Stoytchev}}, \ and\ \bibinfo {author} {\bibfnamefont {H.}~\bibnamefont
  {Safar}},\ }\href@noop {} {\bibfield  {journal} {\bibinfo  {journal} {Phys.
  Today}\ }\textbf {\bibinfo {volume} {54}},\ \bibinfo {pages} {38} (\bibinfo
  {year} {2001})}\BibitemShut {NoStop}%
\bibitem [{\citenamefont {Moustakas}\ \emph {et~al.}(2000)\citenamefont
  {Moustakas}, \citenamefont {Baranger}, \citenamefont {Balents}, \citenamefont
  {Sengupta},\ and\ \citenamefont {Simon}}]{MIMO_Science}%
  \BibitemOpen
  \bibfield  {author} {\bibinfo {author} {\bibfnamefont {A.~L.}\ \bibnamefont
  {Moustakas}}, \bibinfo {author} {\bibfnamefont {H.~U.}\ \bibnamefont
  {Baranger}}, \bibinfo {author} {\bibfnamefont {L.}~\bibnamefont {Balents}},
  \bibinfo {author} {\bibfnamefont {A.~M.}\ \bibnamefont {Sengupta}}, \ and\
  \bibinfo {author} {\bibfnamefont {S.~H.}\ \bibnamefont {Simon}},\ }\href@noop
  {} {\bibfield  {journal} {\bibinfo  {journal} {Science}\ }\textbf {\bibinfo
  {volume} {287}},\ \bibinfo {pages} {287} (\bibinfo {year}
  {2000})}\BibitemShut {NoStop}%
\bibitem [{\citenamefont {{Fink}}(1997)}]{TR_fink}%
  \BibitemOpen
  \bibfield  {author} {\bibinfo {author} {\bibfnamefont {M.}~\bibnamefont
  {{Fink}}},\ }\href {\doibase 10.1063/1.881692} {\bibfield  {journal}
  {\bibinfo  {journal} {Phys. Today}\ }\textbf {\bibinfo {volume} {50}},\
  \bibinfo {pages} {34} (\bibinfo {year} {1997})}\BibitemShut {NoStop}%
\bibitem [{\citenamefont {Lerosey}\ \emph {et~al.}(2004)\citenamefont
  {Lerosey}, \citenamefont {De~Rosny}, \citenamefont {Tourin}, \citenamefont
  {Derode}, \citenamefont {Montaldo},\ and\ \citenamefont {Fink}}]{EMTR_prl}%
  \BibitemOpen
  \bibfield  {author} {\bibinfo {author} {\bibfnamefont {G.}~\bibnamefont
  {Lerosey}}, \bibinfo {author} {\bibfnamefont {J.}~\bibnamefont {De~Rosny}},
  \bibinfo {author} {\bibfnamefont {A.}~\bibnamefont {Tourin}}, \bibinfo
  {author} {\bibfnamefont {A.}~\bibnamefont {Derode}}, \bibinfo {author}
  {\bibfnamefont {G.}~\bibnamefont {Montaldo}}, \ and\ \bibinfo {author}
  {\bibfnamefont {M.}~\bibnamefont {Fink}},\ }\href@noop {} {\bibfield
  {journal} {\bibinfo  {journal} {Phys. Rev. Lett.}\ }\textbf {\bibinfo
  {volume} {92}},\ \bibinfo {pages} {193904} (\bibinfo {year}
  {2004})}\BibitemShut {NoStop}%
\bibitem [{\citenamefont {Vellekoop}\ and\ \citenamefont
  {Mosk}(2007)}]{mosk_SLM}%
  \BibitemOpen
  \bibfield  {author} {\bibinfo {author} {\bibfnamefont {I.~M.}\ \bibnamefont
  {Vellekoop}}\ and\ \bibinfo {author} {\bibfnamefont {A.~P.}\ \bibnamefont
  {Mosk}},\ }\href {\doibase 10.1364/OL.32.002309} {\bibfield  {journal}
  {\bibinfo  {journal} {Opt. Lett.}\ }\textbf {\bibinfo {volume} {32}},\
  \bibinfo {pages} {2309} (\bibinfo {year} {2007})}\BibitemShut {NoStop}%
\bibitem [{\citenamefont {Choi}\ \emph {et~al.}(2011)\citenamefont {Choi},
  \citenamefont {Yang}, \citenamefont {Fang-Yen}, \citenamefont {Kang},
  \citenamefont {Lee}, \citenamefont {Dasari}, \citenamefont {Feld},\ and\
  \citenamefont {Choi}}]{choi2011subwavelengthFOC}%
  \BibitemOpen
  \bibfield  {author} {\bibinfo {author} {\bibfnamefont {Y.}~\bibnamefont
  {Choi}}, \bibinfo {author} {\bibfnamefont {T.~D.}\ \bibnamefont {Yang}},
  \bibinfo {author} {\bibfnamefont {C.}~\bibnamefont {Fang-Yen}}, \bibinfo
  {author} {\bibfnamefont {P.}~\bibnamefont {Kang}}, \bibinfo {author}
  {\bibfnamefont {K.~J.}\ \bibnamefont {Lee}}, \bibinfo {author} {\bibfnamefont
  {R.~R.}\ \bibnamefont {Dasari}}, \bibinfo {author} {\bibfnamefont {M.~S.}\
  \bibnamefont {Feld}}, \ and\ \bibinfo {author} {\bibfnamefont
  {W.}~\bibnamefont {Choi}},\ }\href@noop {} {\bibfield  {journal} {\bibinfo
  {journal} {Phys. Rev. Lett.}\ }\textbf {\bibinfo {volume} {107}},\ \bibinfo
  {pages} {023902} (\bibinfo {year} {2011})}\BibitemShut {NoStop}%
\bibitem [{\citenamefont {van Putten}\ \emph {et~al.}(2011)\citenamefont {van
  Putten}, \citenamefont {Akbulut}, \citenamefont {Bertolotti}, \citenamefont
  {Vos}, \citenamefont {Lagendijk},\ and\ \citenamefont
  {Mosk}}]{Mosk_vis_subwavelength_foc_byWS}%
  \BibitemOpen
  \bibfield  {author} {\bibinfo {author} {\bibfnamefont {E.~G.}\ \bibnamefont
  {van Putten}}, \bibinfo {author} {\bibfnamefont {D.}~\bibnamefont {Akbulut}},
  \bibinfo {author} {\bibfnamefont {J.}~\bibnamefont {Bertolotti}}, \bibinfo
  {author} {\bibfnamefont {W.~L.}\ \bibnamefont {Vos}}, \bibinfo {author}
  {\bibfnamefont {A.}~\bibnamefont {Lagendijk}}, \ and\ \bibinfo {author}
  {\bibfnamefont {A.~P.}\ \bibnamefont {Mosk}},\ }\href {\doibase
  10.1103/PhysRevLett.106.193905} {\bibfield  {journal} {\bibinfo  {journal}
  {Phys. Rev. Lett.}\ }\textbf {\bibinfo {volume} {106}},\ \bibinfo {pages}
  {193905} (\bibinfo {year} {2011})}\BibitemShut {NoStop}%
\bibitem [{\citenamefont {Park}\ \emph {et~al.}(2013)\citenamefont {Park},
  \citenamefont {Park}, \citenamefont {Yu}, \citenamefont {Park}, \citenamefont
  {Han}, \citenamefont {Shin}, \citenamefont {Ko}, \citenamefont {Nam},
  \citenamefont {Cho},\ and\ \citenamefont {Park}}]{park2013subwavelength}%
  \BibitemOpen
  \bibfield  {author} {\bibinfo {author} {\bibfnamefont {J.-H.}\ \bibnamefont
  {Park}}, \bibinfo {author} {\bibfnamefont {C.}~\bibnamefont {Park}}, \bibinfo
  {author} {\bibfnamefont {H.}~\bibnamefont {Yu}}, \bibinfo {author}
  {\bibfnamefont {J.}~\bibnamefont {Park}}, \bibinfo {author} {\bibfnamefont
  {S.}~\bibnamefont {Han}}, \bibinfo {author} {\bibfnamefont {J.}~\bibnamefont
  {Shin}}, \bibinfo {author} {\bibfnamefont {S.~H.}\ \bibnamefont {Ko}},
  \bibinfo {author} {\bibfnamefont {K.~T.}\ \bibnamefont {Nam}}, \bibinfo
  {author} {\bibfnamefont {Y.-H.}\ \bibnamefont {Cho}}, \ and\ \bibinfo
  {author} {\bibfnamefont {Y.}~\bibnamefont {Park}},\ }\href@noop {} {\bibfield
   {journal} {\bibinfo  {journal} {Nat. Photonics}\ }\textbf {\bibinfo {volume}
  {7}},\ \bibinfo {pages} {454} (\bibinfo {year} {2013})}\BibitemShut {NoStop}%
\bibitem [{\citenamefont {Vellekoop}\ \emph {et~al.}(2010)\citenamefont
  {Vellekoop}, \citenamefont {Lagendijk},\ and\ \citenamefont
  {Mosk}}]{mosk_disorder4perfectFOC}%
  \BibitemOpen
  \bibfield  {author} {\bibinfo {author} {\bibfnamefont {I.}~\bibnamefont
  {Vellekoop}}, \bibinfo {author} {\bibfnamefont {A.}~\bibnamefont
  {Lagendijk}}, \ and\ \bibinfo {author} {\bibfnamefont {A.}~\bibnamefont
  {Mosk}},\ }\href@noop {} {\bibfield  {journal} {\bibinfo  {journal} {Nat.
  Photonics}\ }\textbf {\bibinfo {volume} {4}},\ \bibinfo {pages} {320}
  (\bibinfo {year} {2010})}\BibitemShut {NoStop}%
\bibitem [{\citenamefont {Aulbach}\ \emph {et~al.}(2011)\citenamefont
  {Aulbach}, \citenamefont {Gjonaj}, \citenamefont {Johnson}, \citenamefont
  {Mosk},\ and\ \citenamefont {Lagendijk}}]{aulbach_STF}%
  \BibitemOpen
  \bibfield  {author} {\bibinfo {author} {\bibfnamefont {J.}~\bibnamefont
  {Aulbach}}, \bibinfo {author} {\bibfnamefont {B.}~\bibnamefont {Gjonaj}},
  \bibinfo {author} {\bibfnamefont {P.~M.}\ \bibnamefont {Johnson}}, \bibinfo
  {author} {\bibfnamefont {A.~P.}\ \bibnamefont {Mosk}}, \ and\ \bibinfo
  {author} {\bibfnamefont {A.}~\bibnamefont {Lagendijk}},\ }\href {\doibase
  10.1103/PhysRevLett.106.103901} {\bibfield  {journal} {\bibinfo  {journal}
  {Phys. Rev. Lett.}\ }\textbf {\bibinfo {volume} {106}},\ \bibinfo {pages}
  {103901} (\bibinfo {year} {2011})}\BibitemShut {NoStop}%
\bibitem [{\citenamefont {Katz}\ \emph {et~al.}(2011)\citenamefont {Katz},
  \citenamefont {Small}, \citenamefont {Bromberg},\ and\ \citenamefont
  {Silberberg}}]{katz_STF}%
  \BibitemOpen
  \bibfield  {author} {\bibinfo {author} {\bibfnamefont {O.}~\bibnamefont
  {Katz}}, \bibinfo {author} {\bibfnamefont {E.}~\bibnamefont {Small}},
  \bibinfo {author} {\bibfnamefont {Y.}~\bibnamefont {Bromberg}}, \ and\
  \bibinfo {author} {\bibfnamefont {Y.}~\bibnamefont {Silberberg}},\
  }\href@noop {} {\bibfield  {journal} {\bibinfo  {journal} {Nat. Photonics}\
  }\textbf {\bibinfo {volume} {5}},\ \bibinfo {pages} {372} (\bibinfo {year}
  {2011})}\BibitemShut {NoStop}%
\bibitem [{\citenamefont {del Hougne}\ \emph
  {et~al.}(2016{\natexlab{a}})\citenamefont {del Hougne}, \citenamefont
  {Lemoult}, \citenamefont {Fink},\ and\ \citenamefont
  {Lerosey}}]{publikation3}%
  \BibitemOpen
  \bibfield  {author} {\bibinfo {author} {\bibfnamefont {P.}~\bibnamefont {del
  Hougne}}, \bibinfo {author} {\bibfnamefont {F.}~\bibnamefont {Lemoult}},
  \bibinfo {author} {\bibfnamefont {M.}~\bibnamefont {Fink}}, \ and\ \bibinfo
  {author} {\bibfnamefont {G.}~\bibnamefont {Lerosey}},\ }\href@noop {}
  {\bibfield  {journal} {\bibinfo  {journal} {Phys. Rev. Lett.}\ }\textbf
  {\bibinfo {volume} {117}},\ \bibinfo {pages} {134302} (\bibinfo {year}
  {2016}{\natexlab{a}})}\BibitemShut {NoStop}%
\bibitem [{\citenamefont {Liutkus}\ \emph {et~al.}(2014)\citenamefont
  {Liutkus}, \citenamefont {Martina}, \citenamefont {Popoff}, \citenamefont
  {Chardon}, \citenamefont {Katz}, \citenamefont {Lerosey}, \citenamefont
  {Gigan}, \citenamefont {Daudet},\ and\ \citenamefont {Carron}}]{liutkus14}%
  \BibitemOpen
  \bibfield  {author} {\bibinfo {author} {\bibfnamefont {A.}~\bibnamefont
  {Liutkus}}, \bibinfo {author} {\bibfnamefont {D.}~\bibnamefont {Martina}},
  \bibinfo {author} {\bibfnamefont {S.}~\bibnamefont {Popoff}}, \bibinfo
  {author} {\bibfnamefont {G.}~\bibnamefont {Chardon}}, \bibinfo {author}
  {\bibfnamefont {O.}~\bibnamefont {Katz}}, \bibinfo {author} {\bibfnamefont
  {G.}~\bibnamefont {Lerosey}}, \bibinfo {author} {\bibfnamefont
  {S.}~\bibnamefont {Gigan}}, \bibinfo {author} {\bibfnamefont
  {L.}~\bibnamefont {Daudet}}, \ and\ \bibinfo {author} {\bibfnamefont
  {I.}~\bibnamefont {Carron}},\ }\href@noop {} {\bibfield  {journal} {\bibinfo
  {journal} {Sci. Rep.}\ }\textbf {\bibinfo {volume} {4}},\ \bibinfo {pages}
  {5552} (\bibinfo {year} {2014})}\BibitemShut {NoStop}%
\bibitem [{\citenamefont {Popoff}\ \emph
  {et~al.}(2010{\natexlab{a}})\citenamefont {Popoff}, \citenamefont {Lerosey},
  \citenamefont {Carminati}, \citenamefont {Fink}, \citenamefont {Boccara},\
  and\ \citenamefont {Gigan}}]{popoff_prl}%
  \BibitemOpen
  \bibfield  {author} {\bibinfo {author} {\bibfnamefont {S.}~\bibnamefont
  {Popoff}}, \bibinfo {author} {\bibfnamefont {G.}~\bibnamefont {Lerosey}},
  \bibinfo {author} {\bibfnamefont {R.}~\bibnamefont {Carminati}}, \bibinfo
  {author} {\bibfnamefont {M.}~\bibnamefont {Fink}}, \bibinfo {author}
  {\bibfnamefont {A.}~\bibnamefont {Boccara}}, \ and\ \bibinfo {author}
  {\bibfnamefont {S.}~\bibnamefont {Gigan}},\ }\href@noop {} {\bibfield
  {journal} {\bibinfo  {journal} {Phys. Rev. Lett.}\ }\textbf {\bibinfo
  {volume} {104}},\ \bibinfo {pages} {100601} (\bibinfo {year}
  {2010}{\natexlab{a}})}\BibitemShut {NoStop}%
\bibitem [{\citenamefont {Popoff}\ \emph
  {et~al.}(2010{\natexlab{b}})\citenamefont {Popoff}, \citenamefont {Lerosey},
  \citenamefont {Fink}, \citenamefont {Boccara},\ and\ \citenamefont
  {Gigan}}]{popoff_NatComm}%
  \BibitemOpen
  \bibfield  {author} {\bibinfo {author} {\bibfnamefont {S.}~\bibnamefont
  {Popoff}}, \bibinfo {author} {\bibfnamefont {G.}~\bibnamefont {Lerosey}},
  \bibinfo {author} {\bibfnamefont {M.}~\bibnamefont {Fink}}, \bibinfo {author}
  {\bibfnamefont {A.~C.}\ \bibnamefont {Boccara}}, \ and\ \bibinfo {author}
  {\bibfnamefont {S.}~\bibnamefont {Gigan}},\ }\href@noop {} {\bibfield
  {journal} {\bibinfo  {journal} {Nat. Commun.}\ }\textbf {\bibinfo {volume}
  {1}},\ \bibinfo {pages} {81} (\bibinfo {year}
  {2010}{\natexlab{b}})}\BibitemShut {NoStop}%
\bibitem [{\citenamefont {Dr{\'e}meau}\ \emph {et~al.}(2015)\citenamefont
  {Dr{\'e}meau}, \citenamefont {Liutkus}, \citenamefont {Martina},
  \citenamefont {Katz}, \citenamefont {Sch{\"u}lke}, \citenamefont {Krzakala},
  \citenamefont {Gigan},\ and\ \citenamefont {Daudet}}]{LD_binDMD}%
  \BibitemOpen
  \bibfield  {author} {\bibinfo {author} {\bibfnamefont {A.}~\bibnamefont
  {Dr{\'e}meau}}, \bibinfo {author} {\bibfnamefont {A.}~\bibnamefont
  {Liutkus}}, \bibinfo {author} {\bibfnamefont {D.}~\bibnamefont {Martina}},
  \bibinfo {author} {\bibfnamefont {O.}~\bibnamefont {Katz}}, \bibinfo {author}
  {\bibfnamefont {C.}~\bibnamefont {Sch{\"u}lke}}, \bibinfo {author}
  {\bibfnamefont {F.}~\bibnamefont {Krzakala}}, \bibinfo {author}
  {\bibfnamefont {S.}~\bibnamefont {Gigan}}, \ and\ \bibinfo {author}
  {\bibfnamefont {L.}~\bibnamefont {Daudet}},\ }\href@noop {} {\bibfield
  {journal} {\bibinfo  {journal} {Opt. Express}\ }\textbf {\bibinfo {volume}
  {23}},\ \bibinfo {pages} {11898} (\bibinfo {year} {2015})}\BibitemShut
  {NoStop}%
\bibitem [{\citenamefont {Yu}\ \emph {et~al.}(2013)\citenamefont {Yu},
  \citenamefont {Hillman}, \citenamefont {Choi}, \citenamefont {Lee},
  \citenamefont {Feld}, \citenamefont {Dasari},\ and\ \citenamefont
  {Park}}]{park2013largeTM}%
  \BibitemOpen
  \bibfield  {author} {\bibinfo {author} {\bibfnamefont {H.}~\bibnamefont
  {Yu}}, \bibinfo {author} {\bibfnamefont {T.~R.}\ \bibnamefont {Hillman}},
  \bibinfo {author} {\bibfnamefont {W.}~\bibnamefont {Choi}}, \bibinfo {author}
  {\bibfnamefont {J.~O.}\ \bibnamefont {Lee}}, \bibinfo {author} {\bibfnamefont
  {M.~S.}\ \bibnamefont {Feld}}, \bibinfo {author} {\bibfnamefont {R.~R.}\
  \bibnamefont {Dasari}}, \ and\ \bibinfo {author} {\bibfnamefont
  {Y.}~\bibnamefont {Park}},\ }\href@noop {} {\bibfield  {journal} {\bibinfo
  {journal} {Phys. Rev. Lett.}\ }\textbf {\bibinfo {volume} {111}},\ \bibinfo
  {pages} {153902} (\bibinfo {year} {2013})}\BibitemShut {NoStop}%
\bibitem [{\citenamefont {Mounaix}\ \emph {et~al.}(2016)\citenamefont
  {Mounaix}, \citenamefont {Andreoli}, \citenamefont {Defienne}, \citenamefont
  {Volpe}, \citenamefont {Katz}, \citenamefont {Gr\'esillon},\ and\
  \citenamefont {Gigan}}]{mikael_STF}%
  \BibitemOpen
  \bibfield  {author} {\bibinfo {author} {\bibfnamefont {M.}~\bibnamefont
  {Mounaix}}, \bibinfo {author} {\bibfnamefont {D.}~\bibnamefont {Andreoli}},
  \bibinfo {author} {\bibfnamefont {H.}~\bibnamefont {Defienne}}, \bibinfo
  {author} {\bibfnamefont {G.}~\bibnamefont {Volpe}}, \bibinfo {author}
  {\bibfnamefont {O.}~\bibnamefont {Katz}}, \bibinfo {author} {\bibfnamefont
  {S.}~\bibnamefont {Gr\'esillon}}, \ and\ \bibinfo {author} {\bibfnamefont
  {S.}~\bibnamefont {Gigan}},\ }\href {\doibase 10.1103/PhysRevLett.116.253901}
  {\bibfield  {journal} {\bibinfo  {journal} {Phys. Rev. Lett.}\ }\textbf
  {\bibinfo {volume} {116}},\ \bibinfo {pages} {253901} (\bibinfo {year}
  {2016})}\BibitemShut {NoStop}%
\bibitem [{\citenamefont {del Hougne}\ \emph
  {et~al.}(2016{\natexlab{b}})\citenamefont {del Hougne}, \citenamefont
  {Rajaei}, \citenamefont {Daudet},\ and\ \citenamefont {Lerosey}}]{TM_RevMed}%
  \BibitemOpen
  \bibfield  {author} {\bibinfo {author} {\bibfnamefont {P.}~\bibnamefont {del
  Hougne}}, \bibinfo {author} {\bibfnamefont {B.}~\bibnamefont {Rajaei}},
  \bibinfo {author} {\bibfnamefont {L.}~\bibnamefont {Daudet}}, \ and\ \bibinfo
  {author} {\bibfnamefont {G.}~\bibnamefont {Lerosey}},\ }\href@noop {}
  {\bibfield  {journal} {\bibinfo  {journal} {Opt. Express}\ }\textbf {\bibinfo
  {volume} {24}},\ \bibinfo {pages} {18631} (\bibinfo {year}
  {2016}{\natexlab{b}})}\BibitemShut {NoStop}%
\bibitem [{\citenamefont {Kim}\ \emph {et~al.}(2012)\citenamefont {Kim},
  \citenamefont {Choi}, \citenamefont {Yoon}, \citenamefont {Choi},
  \citenamefont {Kim}, \citenamefont {Park},\ and\ \citenamefont
  {Choi}}]{EigChan_Choi_NPhot}%
  \BibitemOpen
  \bibfield  {author} {\bibinfo {author} {\bibfnamefont {M.}~\bibnamefont
  {Kim}}, \bibinfo {author} {\bibfnamefont {Y.}~\bibnamefont {Choi}}, \bibinfo
  {author} {\bibfnamefont {C.}~\bibnamefont {Yoon}}, \bibinfo {author}
  {\bibfnamefont {W.}~\bibnamefont {Choi}}, \bibinfo {author} {\bibfnamefont
  {J.}~\bibnamefont {Kim}}, \bibinfo {author} {\bibfnamefont {Q.-H.}\
  \bibnamefont {Park}}, \ and\ \bibinfo {author} {\bibfnamefont
  {W.}~\bibnamefont {Choi}},\ }\href@noop {} {\bibfield  {journal} {\bibinfo
  {journal} {Nat. Photonics}\ }\textbf {\bibinfo {volume} {6}},\ \bibinfo
  {pages} {581} (\bibinfo {year} {2012})}\BibitemShut {NoStop}%
\bibitem [{\citenamefont {Vellekoop}\ and\ \citenamefont
  {Mosk}(2008{\natexlab{a}})}]{Mosk_WS_OptTrans}%
  \BibitemOpen
  \bibfield  {author} {\bibinfo {author} {\bibfnamefont {I.~M.}\ \bibnamefont
  {Vellekoop}}\ and\ \bibinfo {author} {\bibfnamefont {A.~P.}\ \bibnamefont
  {Mosk}},\ }\href {\doibase 10.1103/PhysRevLett.101.120601} {\bibfield
  {journal} {\bibinfo  {journal} {Phys. Rev. Lett.}\ }\textbf {\bibinfo
  {volume} {101}},\ \bibinfo {pages} {120601} (\bibinfo {year}
  {2008}{\natexlab{a}})}\BibitemShut {NoStop}%
\bibitem [{\citenamefont {Pe{\~n}a}\ \emph {et~al.}(2014)\citenamefont
  {Pe{\~n}a}, \citenamefont {Girschik}, \citenamefont {Libisch}, \citenamefont
  {Rotter},\ and\ \citenamefont {Chabanov}}]{pena2014single}%
  \BibitemOpen
  \bibfield  {author} {\bibinfo {author} {\bibfnamefont {A.}~\bibnamefont
  {Pe{\~n}a}}, \bibinfo {author} {\bibfnamefont {A.}~\bibnamefont {Girschik}},
  \bibinfo {author} {\bibfnamefont {F.}~\bibnamefont {Libisch}}, \bibinfo
  {author} {\bibfnamefont {S.}~\bibnamefont {Rotter}}, \ and\ \bibinfo {author}
  {\bibfnamefont {A.}~\bibnamefont {Chabanov}},\ }\href@noop {} {\bibfield
  {journal} {\bibinfo  {journal} {Nat. Commun.}\ }\textbf {\bibinfo {volume}
  {5}} (\bibinfo {year} {2014})}\BibitemShut {NoStop}%
\bibitem [{\citenamefont {Kaina}\ \emph
  {et~al.}(2014{\natexlab{a}})\citenamefont {Kaina}, \citenamefont {Dupr{\'e}},
  \citenamefont {Lerosey},\ and\ \citenamefont {Fink}}]{SMM_PoC}%
  \BibitemOpen
  \bibfield  {author} {\bibinfo {author} {\bibfnamefont {N.}~\bibnamefont
  {Kaina}}, \bibinfo {author} {\bibfnamefont {M.}~\bibnamefont {Dupr{\'e}}},
  \bibinfo {author} {\bibfnamefont {G.}~\bibnamefont {Lerosey}}, \ and\
  \bibinfo {author} {\bibfnamefont {M.}~\bibnamefont {Fink}},\ }\href@noop {}
  {\bibfield  {journal} {\bibinfo  {journal} {Sci. Rep.}\ }\textbf {\bibinfo
  {volume} {4}} (\bibinfo {year} {2014}{\natexlab{a}})}\BibitemShut {NoStop}%
\bibitem [{\citenamefont {Vellekoop}\ \emph {et~al.}(2008)\citenamefont
  {Vellekoop}, \citenamefont {Van~Putten}, \citenamefont {Lagendijk},\ and\
  \citenamefont {Mosk}}]{FluoMosk}%
  \BibitemOpen
  \bibfield  {author} {\bibinfo {author} {\bibfnamefont {I.}~\bibnamefont
  {Vellekoop}}, \bibinfo {author} {\bibfnamefont {E.}~\bibnamefont
  {Van~Putten}}, \bibinfo {author} {\bibfnamefont {A.}~\bibnamefont
  {Lagendijk}}, \ and\ \bibinfo {author} {\bibfnamefont {A.}~\bibnamefont
  {Mosk}},\ }\href@noop {} {\bibfield  {journal} {\bibinfo  {journal} {Opt.
  Express}\ }\textbf {\bibinfo {volume} {16}},\ \bibinfo {pages} {67} (\bibinfo
  {year} {2008})}\BibitemShut {NoStop}%
\bibitem [{\citenamefont {Larrat}\ \emph {et~al.}(2010)\citenamefont {Larrat},
  \citenamefont {Pernot}, \citenamefont {Montaldo}, \citenamefont {Fink},\ and\
  \citenamefont {Tanter}}]{MRfeedback}%
  \BibitemOpen
  \bibfield  {author} {\bibinfo {author} {\bibfnamefont {B.}~\bibnamefont
  {Larrat}}, \bibinfo {author} {\bibfnamefont {M.}~\bibnamefont {Pernot}},
  \bibinfo {author} {\bibfnamefont {G.}~\bibnamefont {Montaldo}}, \bibinfo
  {author} {\bibfnamefont {M.}~\bibnamefont {Fink}}, \ and\ \bibinfo {author}
  {\bibfnamefont {M.}~\bibnamefont {Tanter}},\ }\href@noop {} {\bibfield
  {journal} {\bibinfo  {journal} {IEEE Trans. Ultrason., Ferroelect., Freq.
  Control}\ }\textbf {\bibinfo {volume} {57}},\ \bibinfo {pages} {1734}
  (\bibinfo {year} {2010})}\BibitemShut {NoStop}%
\bibitem [{\citenamefont {Chaigne}\ \emph {et~al.}(2014)\citenamefont
  {Chaigne}, \citenamefont {Katz}, \citenamefont {Boccara}, \citenamefont
  {Fink}, \citenamefont {Bossy},\ and\ \citenamefont
  {Gigan}}]{Gigan_NonInvTM_PA}%
  \BibitemOpen
  \bibfield  {author} {\bibinfo {author} {\bibfnamefont {T.}~\bibnamefont
  {Chaigne}}, \bibinfo {author} {\bibfnamefont {O.}~\bibnamefont {Katz}},
  \bibinfo {author} {\bibfnamefont {A.~C.}\ \bibnamefont {Boccara}}, \bibinfo
  {author} {\bibfnamefont {M.}~\bibnamefont {Fink}}, \bibinfo {author}
  {\bibfnamefont {E.}~\bibnamefont {Bossy}}, \ and\ \bibinfo {author}
  {\bibfnamefont {S.}~\bibnamefont {Gigan}},\ }\href@noop {} {\bibfield
  {journal} {\bibinfo  {journal} {Nat. Photonics}\ }\textbf {\bibinfo {volume}
  {8}},\ \bibinfo {pages} {58} (\bibinfo {year} {2014})}\BibitemShut {NoStop}%
\bibitem [{\citenamefont {Katz}\ \emph {et~al.}(2014)\citenamefont {Katz},
  \citenamefont {Small}, \citenamefont {Guan},\ and\ \citenamefont
  {Silberberg}}]{OriNL}%
  \BibitemOpen
  \bibfield  {author} {\bibinfo {author} {\bibfnamefont {O.}~\bibnamefont
  {Katz}}, \bibinfo {author} {\bibfnamefont {E.}~\bibnamefont {Small}},
  \bibinfo {author} {\bibfnamefont {Y.}~\bibnamefont {Guan}}, \ and\ \bibinfo
  {author} {\bibfnamefont {Y.}~\bibnamefont {Silberberg}},\ }\href@noop {}
  {\bibfield  {journal} {\bibinfo  {journal} {Optica}\ }\textbf {\bibinfo
  {volume} {1}},\ \bibinfo {pages} {170} (\bibinfo {year} {2014})}\BibitemShut
  {NoStop}%
\bibitem [{\citenamefont {Talla}\ \emph {et~al.}(2015)\citenamefont {Talla},
  \citenamefont {Kellogg}, \citenamefont {Ransford}, \citenamefont
  {Naderiparizi}, \citenamefont {Gollakota},\ and\ \citenamefont
  {Smith}}]{PowByWifi}%
  \BibitemOpen
  \bibfield  {author} {\bibinfo {author} {\bibfnamefont {V.}~\bibnamefont
  {Talla}}, \bibinfo {author} {\bibfnamefont {B.}~\bibnamefont {Kellogg}},
  \bibinfo {author} {\bibfnamefont {B.}~\bibnamefont {Ransford}}, \bibinfo
  {author} {\bibfnamefont {S.}~\bibnamefont {Naderiparizi}}, \bibinfo {author}
  {\bibfnamefont {S.}~\bibnamefont {Gollakota}}, \ and\ \bibinfo {author}
  {\bibfnamefont {J.~R.}\ \bibnamefont {Smith}},\ }in\ \href@noop {} {\emph
  {\bibinfo {booktitle} {Proceedings of the 11th ACM Conference on Emerging
  Networking Experiments and Technologies}}}\ (\bibinfo {organization} {ACM},\
  \bibinfo {year} {2015})\ p.~\bibinfo {pages} {4}\BibitemShut {NoStop}%
\bibitem [{\citenamefont {Kaina}\ \emph
  {et~al.}(2014{\natexlab{b}})\citenamefont {Kaina}, \citenamefont {Dupr\'{e}},
  \citenamefont {Fink},\ and\ \citenamefont {Lerosey}}]{SMM_design}%
  \BibitemOpen
  \bibfield  {author} {\bibinfo {author} {\bibfnamefont {N.}~\bibnamefont
  {Kaina}}, \bibinfo {author} {\bibfnamefont {M.}~\bibnamefont {Dupr\'{e}}},
  \bibinfo {author} {\bibfnamefont {M.}~\bibnamefont {Fink}}, \ and\ \bibinfo
  {author} {\bibfnamefont {G.}~\bibnamefont {Lerosey}},\ }\href {\doibase
  10.1364/OE.22.018881} {\bibfield  {journal} {\bibinfo  {journal} {Opt.
  Express}\ }\textbf {\bibinfo {volume} {22}},\ \bibinfo {pages} {18881}
  (\bibinfo {year} {2014}{\natexlab{b}})}\BibitemShut {NoStop}%
\bibitem [{\citenamefont {Hill}(2009)}]{hill_electromagnetic_2009}%
  \BibitemOpen
  \bibfield  {author} {\bibinfo {author} {\bibfnamefont {D.~A.}\ \bibnamefont
  {Hill}},\ }\href@noop {} {\emph {\bibinfo {title} {Electromagnetic Fields in
  Cavities: Deterministic and Statistical Theories}}},\ Vol.~\bibinfo {volume}
  {35}\ (\bibinfo  {publisher} {John Wiley \& Sons},\ \bibinfo {year}
  {2009})\BibitemShut {NoStop}%
\bibitem [{\citenamefont {Gros}\ \emph {et~al.}(2014)\citenamefont {Gros},
  \citenamefont {Kuhl}, \citenamefont {Legrand}, \citenamefont {Mortessagne},
  \citenamefont {Richalot},\ and\ \citenamefont {Savin}}]{JB_PRL}%
  \BibitemOpen
  \bibfield  {author} {\bibinfo {author} {\bibfnamefont {J.-B.}\ \bibnamefont
  {Gros}}, \bibinfo {author} {\bibfnamefont {U.}~\bibnamefont {Kuhl}}, \bibinfo
  {author} {\bibfnamefont {O.}~\bibnamefont {Legrand}}, \bibinfo {author}
  {\bibfnamefont {F.}~\bibnamefont {Mortessagne}}, \bibinfo {author}
  {\bibfnamefont {E.}~\bibnamefont {Richalot}}, \ and\ \bibinfo {author}
  {\bibfnamefont {D.}~\bibnamefont {Savin}},\ }\href@noop {} {\bibfield
  {journal} {\bibinfo  {journal} {Phys. Rev. Lett.}\ }\textbf {\bibinfo
  {volume} {113}},\ \bibinfo {pages} {224101} (\bibinfo {year}
  {2014})}\BibitemShut {NoStop}%
\bibitem [{\citenamefont {Gros}\ \emph {et~al.}(2016)\citenamefont {Gros},
  \citenamefont {Kuhl}, \citenamefont {Legrand},\ and\ \citenamefont
  {Mortessagne}}]{JB_PRE}%
  \BibitemOpen
  \bibfield  {author} {\bibinfo {author} {\bibfnamefont {J.-B.}\ \bibnamefont
  {Gros}}, \bibinfo {author} {\bibfnamefont {U.}~\bibnamefont {Kuhl}}, \bibinfo
  {author} {\bibfnamefont {O.}~\bibnamefont {Legrand}}, \ and\ \bibinfo
  {author} {\bibfnamefont {F.}~\bibnamefont {Mortessagne}},\ }\href@noop {}
  {\bibfield  {journal} {\bibinfo  {journal} {Phys. Rev. E}\ }\textbf {\bibinfo
  {volume} {93}},\ \bibinfo {pages} {032108} (\bibinfo {year}
  {2016})}\BibitemShut {NoStop}%
%
\bibitem[{\citenamefont{Fromenteze et~al.}(2015)\citenamefont{Fromenteze,
  Yurduseven, Imani, Gollub, Decroze, Carsenat, and
  Smith}}]{DavidSmith_CompImag_APL}
\bibinfo{author}{\bibfnamefont{T.}~\bibnamefont{Fromenteze}},
  \bibinfo{author}{\bibfnamefont{O.}~\bibnamefont{Yurduseven}},
  \bibinfo{author}{\bibfnamefont{M.~F.} \bibnamefont{Imani}},
  \bibinfo{author}{\bibfnamefont{J.}~\bibnamefont{Gollub}},
  \bibinfo{author}{\bibfnamefont{C.}~\bibnamefont{Decroze}},
  \bibinfo{author}{\bibfnamefont{D.}~\bibnamefont{Carsenat}}, \bibnamefont{and}
  \bibinfo{author}{\bibfnamefont{D.~R.} \bibnamefont{Smith}},
  \bibinfo{journal}{Appl. Phys. Lett.} \textbf{\bibinfo{volume}{106}},
  \bibinfo{eid}{194104} (\bibinfo{year}{2015}).
%
%
\bibitem [{\citenamefont {Sleasman}\ \emph {et~al.}(2016)\citenamefont
  {Sleasman}, \citenamefont {Imani}, \citenamefont {Gollub},\ and\
  \citenamefont {Smith}}]{sleasman2016microwave}%
  \BibitemOpen
  \bibfield  {author} {\bibinfo {author} {\bibfnamefont {T.}~\bibnamefont
  {Sleasman}}, \bibinfo {author} {\bibfnamefont {M.~F.}\ \bibnamefont {Imani}},
  \bibinfo {author} {\bibfnamefont {J.~N.}\ \bibnamefont {Gollub}}, \ and\
  \bibinfo {author} {\bibfnamefont {D.~R.}\ \bibnamefont {Smith}},\ }\href@noop
  {} {\bibfield  {journal} {\bibinfo  {journal} {Phys. Rev. Applied}\ }\textbf
  {\bibinfo {volume} {6}},\ \bibinfo {pages} {054019} (\bibinfo {year}
  {2016})}\BibitemShut {NoStop}%
\bibitem [{\citenamefont {F.~Imani}\ \emph {et~al.}(2016)\citenamefont
  {F.~Imani}, \citenamefont {Sleasman}, \citenamefont {Gollub},\ and\
  \citenamefont {Smith}}]{SmithNumerical}%
  \BibitemOpen
  \bibfield  {author} {\bibinfo {author} {\bibfnamefont {M.}~\bibnamefont
  {F.~Imani}}, \bibinfo {author} {\bibfnamefont {T.}~\bibnamefont {Sleasman}},
  \bibinfo {author} {\bibfnamefont {J.~N.}\ \bibnamefont {Gollub}}, \ and\
  \bibinfo {author} {\bibfnamefont {D.~R.}\ \bibnamefont {Smith}},\ }\href@noop
  {} {\bibfield  {journal} {\bibinfo  {journal} {J. Appl. Phys.}\ }\textbf
  {\bibinfo {volume} {120}},\ \bibinfo {pages} {144903} (\bibinfo {year}
  {2016})}\BibitemShut {NoStop}%
\bibitem [{\citenamefont {Gollub}\ \emph {et~al.}(2017)\citenamefont {Gollub},
  \citenamefont {Yurduseven}, \citenamefont {Trofatter}, \citenamefont
  {Arnitz}, \citenamefont {Imani}, \citenamefont {Sleasman}, \citenamefont
  {Boyarsky}, \citenamefont {Rose}, \citenamefont {Pedross-Engel},
  \citenamefont {Odabasi} \emph {et~al.}}]{davidsmithscirep}%
  \BibitemOpen
  \bibfield  {author} {\bibinfo {author} {\bibfnamefont {J.}~\bibnamefont
  {Gollub}}, \bibinfo {author} {\bibfnamefont {O.}~\bibnamefont {Yurduseven}},
  \bibinfo {author} {\bibfnamefont {K.}~\bibnamefont {Trofatter}}, \bibinfo
  {author} {\bibfnamefont {D.}~\bibnamefont {Arnitz}}, \bibinfo {author}
  {\bibfnamefont {M.}~\bibnamefont {Imani}}, \bibinfo {author} {\bibfnamefont
  {T.}~\bibnamefont {Sleasman}}, \bibinfo {author} {\bibfnamefont
  {M.}~\bibnamefont {Boyarsky}}, \bibinfo {author} {\bibfnamefont
  {A.}~\bibnamefont {Rose}}, \bibinfo {author} {\bibfnamefont {A.}~\bibnamefont
  {Pedross-Engel}}, \bibinfo {author} {\bibfnamefont {H.}~\bibnamefont
  {Odabasi}},  \emph {et~al.},\ }\href@noop {} {\bibfield  {journal} {\bibinfo
  {journal} {Sci. Rep.}\ }\textbf {\bibinfo {volume} {7}} (\bibinfo {year}
  {2017})}\BibitemShut {NoStop}%
\bibitem [{\citenamefont {Haboubi}\ \emph {et~al.}(2014)\citenamefont
  {Haboubi}, \citenamefont {Takhedmit}, \citenamefont {Lan Sun~Luk},
  \citenamefont {Adami}, \citenamefont {Allard}, \citenamefont {Costa},
  \citenamefont {Vollaire}, \citenamefont {Picon},\ and\ \citenamefont
  {Cirio}}]{HarviDesign}%
  \BibitemOpen
  \bibfield  {author} {\bibinfo {author} {\bibfnamefont {W.}~\bibnamefont
  {Haboubi}}, \bibinfo {author} {\bibfnamefont {H.}~\bibnamefont {Takhedmit}},
  \bibinfo {author} {\bibfnamefont {J.-D.}\ \bibnamefont {Lan Sun~Luk}},
  \bibinfo {author} {\bibfnamefont {S.-E.}\ \bibnamefont {Adami}}, \bibinfo
  {author} {\bibfnamefont {B.}~\bibnamefont {Allard}}, \bibinfo {author}
  {\bibfnamefont {F.}~\bibnamefont {Costa}}, \bibinfo {author} {\bibfnamefont
  {C.}~\bibnamefont {Vollaire}}, \bibinfo {author} {\bibfnamefont
  {O.}~\bibnamefont {Picon}}, \ and\ \bibinfo {author} {\bibfnamefont
  {L.}~\bibnamefont {Cirio}},\ }\href@noop {} {\bibfield  {journal} {\bibinfo
  {journal} {Progress In Electromagnetics Research}\ }\textbf {\bibinfo
  {volume} {148}},\ \bibinfo {pages} {31} (\bibinfo {year} {2014})}\BibitemShut
  {NoStop}%
\bibitem [{\citenamefont {Dupr\'e}\ \emph {et~al.}(2015)\citenamefont
  {Dupr\'e}, \citenamefont {del Hougne}, \citenamefont {Fink}, \citenamefont
  {Lemoult},\ and\ \citenamefont {Lerosey}}]{publikation1}%
  \BibitemOpen
  \bibfield  {author} {\bibinfo {author} {\bibfnamefont {M.}~\bibnamefont
  {Dupr\'e}}, \bibinfo {author} {\bibfnamefont {P.}~\bibnamefont {del Hougne}},
  \bibinfo {author} {\bibfnamefont {M.}~\bibnamefont {Fink}}, \bibinfo {author}
  {\bibfnamefont {F.}~\bibnamefont {Lemoult}}, \ and\ \bibinfo {author}
  {\bibfnamefont {G.}~\bibnamefont {Lerosey}},\ }\href {\doibase
  10.1103/PhysRevLett.115.017701} {\bibfield  {journal} {\bibinfo  {journal}
  {Phys. Rev. Lett.}\ }\textbf {\bibinfo {volume} {115}},\ \bibinfo {pages}
  {017701} (\bibinfo {year} {2015})}\BibitemShut {NoStop}%
\bibitem [{\citenamefont {Vellekoop}\ and\ \citenamefont
  {Mosk}(2008{\natexlab{b}})}]{moskWSalgo}%
  \BibitemOpen
  \bibfield  {author} {\bibinfo {author} {\bibfnamefont {I.}~\bibnamefont
  {Vellekoop}}\ and\ \bibinfo {author} {\bibfnamefont {A.}~\bibnamefont
  {Mosk}},\ }\href@noop {} {\bibfield  {journal} {\bibinfo  {journal} {Opt.
  Commun.}\ }\textbf {\bibinfo {volume} {281}},\ \bibinfo {pages} {3071}
  (\bibinfo {year} {2008}{\natexlab{b}})}\BibitemShut {NoStop}%
\bibitem [{Note1()}]{Note1}%
  \BibitemOpen
  \bibinfo {note} {Bandpass filtered long series of numbers generated with
  MatLab's uniform random number generator.}\BibitemShut {Stop}%
\bibitem [{\citenamefont {Paudel}\ \emph {et~al.}(2013)\citenamefont {Paudel},
  \citenamefont {Stockbridge}, \citenamefont {Mertz},\ and\ \citenamefont
  {Bifano}}]{SpectIncoh1}%
  \BibitemOpen
  \bibfield  {author} {\bibinfo {author} {\bibfnamefont {H.~P.}\ \bibnamefont
  {Paudel}}, \bibinfo {author} {\bibfnamefont {C.}~\bibnamefont {Stockbridge}},
  \bibinfo {author} {\bibfnamefont {J.}~\bibnamefont {Mertz}}, \ and\ \bibinfo
  {author} {\bibfnamefont {T.}~\bibnamefont {Bifano}},\ }\href@noop {}
  {\bibfield  {journal} {\bibinfo  {journal} {Opt. Express}\ }\textbf {\bibinfo
  {volume} {21}},\ \bibinfo {pages} {17299} (\bibinfo {year}
  {2013})}\BibitemShut {NoStop}%
\bibitem [{\citenamefont {Andreoli}\ \emph {et~al.}(2015)\citenamefont
  {Andreoli}, \citenamefont {Volpe}, \citenamefont {Popoff}, \citenamefont
  {Katz}, \citenamefont {Gr{\'e}sillon},\ and\ \citenamefont
  {Gigan}}]{SpectIncoh2}%
  \BibitemOpen
  \bibfield  {author} {\bibinfo {author} {\bibfnamefont {D.}~\bibnamefont
  {Andreoli}}, \bibinfo {author} {\bibfnamefont {G.}~\bibnamefont {Volpe}},
  \bibinfo {author} {\bibfnamefont {S.}~\bibnamefont {Popoff}}, \bibinfo
  {author} {\bibfnamefont {O.}~\bibnamefont {Katz}}, \bibinfo {author}
  {\bibfnamefont {S.}~\bibnamefont {Gr{\'e}sillon}}, \ and\ \bibinfo {author}
  {\bibfnamefont {S.}~\bibnamefont {Gigan}},\ }\href@noop {} {\bibfield
  {journal} {\bibinfo  {journal} {Sci. Rep.}\ }\textbf {\bibinfo {volume}
  {5}},\ \bibinfo {pages} {10347} (\bibinfo {year} {2015})}\BibitemShut
  {NoStop}%
\bibitem [{\citenamefont {Small}\ \emph {et~al.}(2012)\citenamefont {Small},
  \citenamefont {Katz}, \citenamefont {Guan},\ and\ \citenamefont
  {Silberberg}}]{SpectIncoh3}%
  \BibitemOpen
  \bibfield  {author} {\bibinfo {author} {\bibfnamefont {E.}~\bibnamefont
  {Small}}, \bibinfo {author} {\bibfnamefont {O.}~\bibnamefont {Katz}},
  \bibinfo {author} {\bibfnamefont {Y.}~\bibnamefont {Guan}}, \ and\ \bibinfo
  {author} {\bibfnamefont {Y.}~\bibnamefont {Silberberg}},\ }\href@noop {}
  {\bibfield  {journal} {\bibinfo  {journal} {Opt. Lett.}\ }\textbf {\bibinfo
  {volume} {37}},\ \bibinfo {pages} {3429} (\bibinfo {year}
  {2012})}\BibitemShut {NoStop}%
\bibitem [{\citenamefont {Van~Beijnum}\ \emph {et~al.}(2011)\citenamefont
  {Van~Beijnum}, \citenamefont {Van~Putten}, \citenamefont {Lagendijk},\ and\
  \citenamefont {Mosk}}]{SpectIncoh4}%
  \BibitemOpen
  \bibfield  {author} {\bibinfo {author} {\bibfnamefont {F.}~\bibnamefont
  {Van~Beijnum}}, \bibinfo {author} {\bibfnamefont {E.~G.}\ \bibnamefont
  {Van~Putten}}, \bibinfo {author} {\bibfnamefont {A.}~\bibnamefont
  {Lagendijk}}, \ and\ \bibinfo {author} {\bibfnamefont {A.~P.}\ \bibnamefont
  {Mosk}},\ }\href@noop {} {\bibfield  {journal} {\bibinfo  {journal} {Opt.
  Lett.}\ }\textbf {\bibinfo {volume} {36}},\ \bibinfo {pages} {373} (\bibinfo
  {year} {2011})}\BibitemShut {NoStop}%
\bibitem [{\citenamefont {Hsu}\ \emph {et~al.}(2015)\citenamefont {Hsu},
  \citenamefont {Goetschy}, \citenamefont {Bromberg}, \citenamefont {Stone},\
  and\ \citenamefont {Cao}}]{SpectIncoh5}%
  \BibitemOpen
  \bibfield  {author} {\bibinfo {author} {\bibfnamefont {C.~W.}\ \bibnamefont
  {Hsu}}, \bibinfo {author} {\bibfnamefont {A.}~\bibnamefont {Goetschy}},
  \bibinfo {author} {\bibfnamefont {Y.}~\bibnamefont {Bromberg}}, \bibinfo
  {author} {\bibfnamefont {A.~D.}\ \bibnamefont {Stone}}, \ and\ \bibinfo
  {author} {\bibfnamefont {H.}~\bibnamefont {Cao}},\ }\href@noop {} {\bibfield
  {journal} {\bibinfo  {journal} {Phys. Rev. Lett.}\ }\textbf {\bibinfo
  {volume} {115}},\ \bibinfo {pages} {223901} (\bibinfo {year}
  {2015})}\BibitemShut {NoStop}%
\bibitem [{\citenamefont {Frazier}\ \emph
  {et~al.}(2013{\natexlab{a}})\citenamefont {Frazier}, \citenamefont {Taddese},
  \citenamefont {Xiao}, \citenamefont {Antonsen}, \citenamefont {Ott},\ and\
  \citenamefont {Anlage}}]{AnlageNLTR}%
  \BibitemOpen
  \bibfield  {author} {\bibinfo {author} {\bibfnamefont {M.}~\bibnamefont
  {Frazier}}, \bibinfo {author} {\bibfnamefont {B.}~\bibnamefont {Taddese}},
  \bibinfo {author} {\bibfnamefont {B.}~\bibnamefont {Xiao}}, \bibinfo {author}
  {\bibfnamefont {T.}~\bibnamefont {Antonsen}}, \bibinfo {author}
  {\bibfnamefont {E.}~\bibnamefont {Ott}}, \ and\ \bibinfo {author}
  {\bibfnamefont {S.~M.}\ \bibnamefont {Anlage}},\ }\href@noop {} {\bibfield
  {journal} {\bibinfo  {journal} {Phys. Rev. E}\ }\textbf {\bibinfo {volume}
  {88}},\ \bibinfo {pages} {062910} (\bibinfo {year}
  {2013}{\natexlab{a}})}\BibitemShut {NoStop}%
\bibitem [{\citenamefont {Frazier}\ \emph
  {et~al.}(2013{\natexlab{b}})\citenamefont {Frazier}, \citenamefont {Taddese},
  \citenamefont {Antonsen},\ and\ \citenamefont {Anlage}}]{AnlageNLprl}%
  \BibitemOpen
  \bibfield  {author} {\bibinfo {author} {\bibfnamefont {M.}~\bibnamefont
  {Frazier}}, \bibinfo {author} {\bibfnamefont {B.}~\bibnamefont {Taddese}},
  \bibinfo {author} {\bibfnamefont {T.}~\bibnamefont {Antonsen}}, \ and\
  \bibinfo {author} {\bibfnamefont {S.~M.}\ \bibnamefont {Anlage}},\
  }\href@noop {} {\bibfield  {journal} {\bibinfo  {journal} {Phys. Rev. Lett.}\
  }\textbf {\bibinfo {volume} {110}},\ \bibinfo {pages} {063902} (\bibinfo
  {year} {2013}{\natexlab{b}})}\BibitemShut {NoStop}%
\bibitem [{\citenamefont {Hand}\ and\ \citenamefont {Cummer}(2010)}]{RAcummer}%
  \BibitemOpen
  \bibfield  {author} {\bibinfo {author} {\bibfnamefont {T.~H.}\ \bibnamefont
  {Hand}}\ and\ \bibinfo {author} {\bibfnamefont {S.~A.}\ \bibnamefont
  {Cummer}},\ }\href@noop {} {\bibfield  {journal} {\bibinfo  {journal} {IEEE
  Antennas Wireless Propag. Lett.}\ }\textbf {\bibinfo {volume} {9}},\ \bibinfo
  {pages} {70} (\bibinfo {year} {2010})}\BibitemShut {NoStop}%
\bibitem [{\citenamefont {Hum}\ and\ \citenamefont
  {Perruisseau-Carrier}(2014)}]{RApc}%
  \BibitemOpen
  \bibfield  {author} {\bibinfo {author} {\bibfnamefont {S.~V.}\ \bibnamefont
  {Hum}}\ and\ \bibinfo {author} {\bibfnamefont {J.}~\bibnamefont
  {Perruisseau-Carrier}},\ }\href@noop {} {\bibfield  {journal} {\bibinfo
  {journal} {IEEE Trans. Antennas Propag.}\ }\textbf {\bibinfo {volume} {62}},\
  \bibinfo {pages} {183} (\bibinfo {year} {2014})}\BibitemShut {NoStop}%
\bibitem [{\citenamefont {Yang}\ \emph {et~al.}(2016)\citenamefont {Yang},
  \citenamefont {Cao}, \citenamefont {Yang}, \citenamefont {Gao}, \citenamefont
  {Xu}, \citenamefont {Li}, \citenamefont {Chen}, \citenamefont {Zhao},
  \citenamefont {Zheng},\ and\ \citenamefont {Li}}]{RAli}%
  \BibitemOpen
  \bibfield  {author} {\bibinfo {author} {\bibfnamefont {H.}~\bibnamefont
  {Yang}}, \bibinfo {author} {\bibfnamefont {X.}~\bibnamefont {Cao}}, \bibinfo
  {author} {\bibfnamefont {F.}~\bibnamefont {Yang}}, \bibinfo {author}
  {\bibfnamefont {J.}~\bibnamefont {Gao}}, \bibinfo {author} {\bibfnamefont
  {S.}~\bibnamefont {Xu}}, \bibinfo {author} {\bibfnamefont {M.}~\bibnamefont
  {Li}}, \bibinfo {author} {\bibfnamefont {X.}~\bibnamefont {Chen}}, \bibinfo
  {author} {\bibfnamefont {Y.}~\bibnamefont {Zhao}}, \bibinfo {author}
  {\bibfnamefont {Y.}~\bibnamefont {Zheng}}, \ and\ \bibinfo {author}
  {\bibfnamefont {S.}~\bibnamefont {Li}},\ }\href@noop {} {\bibfield  {journal}
  {\bibinfo  {journal} {Sci. Rep.}\ }\textbf {\bibinfo {volume} {6}},\ \bibinfo
  {pages} {35692} (\bibinfo {year} {2016})}\BibitemShut {NoStop}%
\bibitem [{\citenamefont {Aulbach}\ \emph {et~al.}(2012)\citenamefont
  {Aulbach}, \citenamefont {Bretagne}, \citenamefont {Fink}, \citenamefont
  {Tanter},\ and\ \citenamefont {Tourin}}]{BretagnePRE}%
  \BibitemOpen
  \bibfield  {author} {\bibinfo {author} {\bibfnamefont {J.}~\bibnamefont
  {Aulbach}}, \bibinfo {author} {\bibfnamefont {A.}~\bibnamefont {Bretagne}},
  \bibinfo {author} {\bibfnamefont {M.}~\bibnamefont {Fink}}, \bibinfo {author}
  {\bibfnamefont {M.}~\bibnamefont {Tanter}}, \ and\ \bibinfo {author}
  {\bibfnamefont {A.}~\bibnamefont {Tourin}},\ }\href@noop {} {\bibfield
  {journal} {\bibinfo  {journal} {Phys. Rev. E}\ }\textbf {\bibinfo {volume}
  {85}},\ \bibinfo {pages} {016605} (\bibinfo {year} {2012})}\BibitemShut
  {NoStop}%
\bibitem [{\citenamefont {Smith}\ \emph {et~al.}(2017)\citenamefont {Smith},
  \citenamefont {Gowda}, \citenamefont {Yurduseven}, \citenamefont {Larouche},
  \citenamefont {Lipworth}, \citenamefont {Urzhumov},\ and\ \citenamefont
  {Reynolds}}]{WTP_Smith}%
  \BibitemOpen
  \bibfield  {author} {\bibinfo {author} {\bibfnamefont {D.~R.}\ \bibnamefont
  {Smith}}, \bibinfo {author} {\bibfnamefont {V.~R.}\ \bibnamefont {Gowda}},
  \bibinfo {author} {\bibfnamefont {O.}~\bibnamefont {Yurduseven}}, \bibinfo
  {author} {\bibfnamefont {S.}~\bibnamefont {Larouche}}, \bibinfo {author}
  {\bibfnamefont {G.}~\bibnamefont {Lipworth}}, \bibinfo {author}
  {\bibfnamefont {Y.}~\bibnamefont {Urzhumov}}, \ and\ \bibinfo {author}
  {\bibfnamefont {M.~S.}\ \bibnamefont {Reynolds}},\ }\href@noop {} {\bibfield
  {journal} {\bibinfo  {journal} {J. Appl. Phys.}\ }\textbf {\bibinfo {volume}
  {121}},\ \bibinfo {pages} {014901} (\bibinfo {year} {2017})}\BibitemShut
  {NoStop}%
\bibitem [{\citenamefont {Xiao}\ \emph {et~al.}(2016)\citenamefont {Xiao},
  \citenamefont {Antonsen}, \citenamefont {Ott},\ and\ \citenamefont
  {Anlage}}]{Anlage_syntheticTR}%
  \BibitemOpen
  \bibfield  {author} {\bibinfo {author} {\bibfnamefont {B.}~\bibnamefont
  {Xiao}}, \bibinfo {author} {\bibfnamefont {T.~M.}\ \bibnamefont {Antonsen}},
  \bibinfo {author} {\bibfnamefont {E.}~\bibnamefont {Ott}}, \ and\ \bibinfo
  {author} {\bibfnamefont {S.~M.}\ \bibnamefont {Anlage}},\ }\href {\doibase
  10.1103/PhysRevE.93.052205} {\bibfield  {journal} {\bibinfo  {journal} {Phys.
  Rev. E}\ }\textbf {\bibinfo {volume} {93}},\ \bibinfo {pages} {052205}
  (\bibinfo {year} {2016})}\BibitemShut {NoStop}%
\bibitem [{\citenamefont {Hong}\ \emph {et~al.}(2014)\citenamefont {Hong},
  \citenamefont {Mendez}, \citenamefont {Koch}, \citenamefont {Wall},\ and\
  \citenamefont {Anlage}}]{AnlageNLapplied}%
  \BibitemOpen
  \bibfield  {author} {\bibinfo {author} {\bibfnamefont {S.~K.}\ \bibnamefont
  {Hong}}, \bibinfo {author} {\bibfnamefont {V.~M.}\ \bibnamefont {Mendez}},
  \bibinfo {author} {\bibfnamefont {T.}~\bibnamefont {Koch}}, \bibinfo {author}
  {\bibfnamefont {W.~S.}\ \bibnamefont {Wall}}, \ and\ \bibinfo {author}
  {\bibfnamefont {S.~M.}\ \bibnamefont {Anlage}},\ }\href@noop {} {\bibfield
  {journal} {\bibinfo  {journal} {Phys. Rev. Applied}\ }\textbf {\bibinfo
  {volume} {2}},\ \bibinfo {pages} {044013} (\bibinfo {year}
  {2014})}\BibitemShut {NoStop}%
\end{thebibliography}
%

\end{document}